\documentclass[
reprint,
amsmath,amssymb,
aps,nofootinbib
]{revtex4-2}

\usepackage{subfig}
\usepackage{graphicx}
\usepackage{dcolumn}
\usepackage{bm}
\usepackage{graphicx}	
\usepackage{amsmath}	
\usepackage{mathtools}  
\usepackage{xcolor}
\usepackage{hyperref}
\usepackage{comment}
\usepackage{tabularx}

\usepackage{caption}
\usepackage{academicons}

\newcommand{\cg}[1]{\textcolor{black}{#1}}

\newcommand{\orcidicon}{\includegraphics[width=0.32cm]{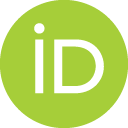}}
\newcommand{\orcid}[1]{\href{https://orcid.org/#1}{\orcidicon}}

\begin{document}
	\title[Dark Matter Minihalo Disruption]{Disruption of Dark Matter Minihalos by Successive  Stellar Encounters}
	
	\author{Ian DSouza\orcid{0009-0004-3142-3898}}
	\email{ids29@uclive.ac.nz}
	\author{Chris Gordon\orcid{0000-0003-4864-5150}}
	\email{chris.gordon@canterbury.ac.nz}
	\affiliation{
		School of Physical and Chemical Sciences, University of Canterbury,
		New Zealand
	}

	\date{\today}
	
	\begin{abstract}
		Scenarios such as the QCD axion with the Peccei-Quinn symmetry broken after inflation predict an enhanced matter power spectrum on sub-parsec scales. These theories lead to the formation of dense dark matter structures known as minihalos, which provide insights into early Universe dynamics and have implications for direct detection experiments. We examine the mass loss of minihalos during stellar encounters, building on previous studies that derived formulas for mass loss and performed N-body simulations. We propose a new formula for the mass loss that accounts for changes in the minihalo profile after disruption by a passing star. We also investigate the mass loss for multiple stellar encounters.
		We demonstrate that accurately assessing the mass loss in minihalos due to multiple stellar encounters necessitates considering the alterations in the minihalo's binding energy after each encounter, as overlooking this aspect results in a substantial underestimation of the mass loss.
	\end{abstract}

	\maketitle

	\section{Introduction}
	
	Beyond the well-known weakly interacting massive particles (WIMPs) paradigm, there are alternative theories that predict an enhanced matter power spectrum on sub-parsec ($\lesssim 10^{-6}M_\odot$) scales. These theories, which include 
	the quantum chromodynamics  axion  with the Peccei-Quinn symmetry \citep{Peccei:1977hh} broken after inflation \citep[{\em e.g.},][]{Hogan:1988mp,kolb1993axion,Kolb1994nonlinear,Zurek:2006sy},
	early matter domination 
	\citep[{\em e.g.},][]{Erickcek:2011us,Fan:2014zua}
	and vector dark matter models \citep[{\em e.g.},][]{Nelson:2011sf, Graham:2015rva}, lead to the formation of dense dark matter structures, which are known as {\em minihalos\/}.
	Dark matter minihalos, in these theories, originate earlier and are denser, making them much less susceptible to disruption compared to models such as those based on WIMPs, which do not have an enhanced matter power spectrum on sub-parsec ($\lesssim 10^{-6}M_\odot$) scales \citep[{\em e.g.},][]{Ostriker1972, Gnedin1999, Goerdt2007, Zhao2007, Schneider2010}.
	Minihalos are potentially observable in local studies \citep[{\em e.g.},][]{Dror:2019twh, Ramani:2020hdo, Lee:2020wfn} and their presence would also have important implications for direct detection experiments\citep[{\em e.g.},][]{eggemeierAxionMinivoidsImplications2023, OHare2023}.

	When a minihalo encounters a star, energy is injected into the minihalo through tidal interactions~\citep[{\em e.g.},][]{binneyTremaine}.
	Ref.~\cite{K2020} (hereafter referred to as K2021) derived a general formula for the mass loss of a minihalo during a stellar encounter using the phase space distribution function of the dark matter particles. A similar formula, using a wave description, for the mass loss, was derived by ref.~\cite{Dandoy:2022prp}. 
	Ref.~\cite{S2022} (hereafter referred to as S2023) performed N-body simulations of dark matter minihalos undergoing a stellar interaction. They varied the normalized injected energy of the minihalo-star interaction, the concentration parameter, and the virial mass of the minihalo and computed the survival fraction of the minihalo. They also used an empirical response function to fit the numerically simulated data. They found that the formula developed by K2021 provided a reasonable fit for a halo concentration of $c=100$. In this article, we show that K2021's formula does not work so well for other concentrations. We derive a formula that performs better on all the concentrations that S2023 showed detailed results for in their paper. Our new formula uses a sequential stripping approach and accounts for the minihalo profile change after being disrupted by a passing star.
	
	S2023 also investigated the mass loss for multiple stellar encounters. This is important as the minihalos in our galaxy will transverse the Galactic disk many times during the history of the Universe [{\em e.g.}, S2023]. 
	The fractional energy is the ratio of the energy injected into the minihalo divided by the minihalo binding energy. S2023
	assumed that the mass loss from multiple stellar encounters depends on
	the sum of the fractional energies from each encounter.
	In this article, we check this result and show that it does not account for the change in the halo profile after each encounter. Using our formula, we find that the mass loss is significantly more severe \cg{than the method employed by S2023}.

	S2023 estimated that 
	about 60\% of mass in minihalos with an initial mass greater than $10^{-12}M_\odot$ will  
	be retained by minihalos observed at the redshift zero at the solar system location.
	Our results indicate that this is an overestimate of the amount of retained mass.

	The paper is structured as follows: Section \ref{sec:Mass_loss_in_a_minhalo_during_a_stellar_encounter} details the K2021 method and its application across various concentration parameter values. Our proposed sequential stripping model is elaborated in Section \ref{sec:The_sequential_stripping_model}. The dynamics of multiple stellar encounters are discussed in Section \ref{sec:Multiple_stellar_encounters_of_an_NFW_minihalo}, leading to our concluding thoughts in Section \ref{sec:Discussion_and_Conclusions}. For more in-depth technical explanations, readers are directed to the appendices.
	
	\section{{Mass loss in a minihalo during a stellar encounter}}
	\label{sec:Mass_loss_in_a_minhalo_during_a_stellar_encounter}
	
	N-body simulations~\citep[{\em e.g.},][]{Xiao:2021nkb} indicate that the undisrupted minihalos can be fit by the well-known
	spherically symmetric Navarro–Frenk–White (NFW) density profile Ref.~\cite{NFW1997universal}.
	
	Mathematically, the NFW density profile $\rho$ is described as a function of the distance $r$ from the center of the minihalo as:\begin{equation}\label{eqn:NFW as a function of r}
		\rho(r) = \frac{\rho_{\rm s}}{\frac{r}{r_{\rm s}}\left(1+\frac{r}{r_{\rm s}}\right)^2}\,.
	\end{equation}
	The parameters $\rho_{\rm s}$ and $r_{\rm s}$ are two independent parameters of the NFW profile. In addition, the virial radius $r_{\rm vir}$ is approximated as the radius within which the mean density is 200 times the cosmological critical density.
	We will use 
	the following normalized distance from the center of the minihalo
	\begin{equation}\label{eq:x definition}
		x \equiv \frac{r}{r_{\rm vir}}
	\end{equation}
	This is sometimes also called the normalized radius. 
	The concentration parameter of the minihalo is defined as
	\begin{equation}\label{eq:concentration defintion}
		c \equiv \frac{r_{\rm vir}}{r_{\rm s}}\, .
	\end{equation}
	Using these variables, we can rewrite the NFW profile as 
	\begin{equation}\label{eqn:NFW as a function of x}
		\rho(x) = \frac{\rho_{\rm s}}{cx\left(1+cx\right)^2}\,.
	\end{equation}

	Consider a minihalo with an NFW density profile extending to infinity. We are interested in the mass loss during a stellar encounter within the virial radius of the minihalo.
	
	As in S2023, the terms $\langle r^2 \rangle$ and $\langle r^{-2} \rangle$ both averaged within the virial radius are parametrized by $\alpha$ and $\beta$ as follows:
	\begin{align}\label{eq:langle r^2 rangle}
		\langle r^2 \rangle & \equiv \alpha^2 r_{\rm vir}^2\\
		\langle r^{-2} \rangle & \equiv \beta^2 r_{\rm vir}^{\cg{-}2}
	\end{align}
	The binding energy $E_{\rm b}$ of the minihalo within the virial radius is parametrized by $\gamma$ as follows (S2023):
	\begin{equation}\label{eq:binding energy}
		E_{\rm b} = \gamma \frac{GM_{\rm vir}^2}{r_{\rm vir}}
	\end{equation}
	where $M_{\rm vir}$ is the mass of the minihalo contained within the virial radius.
	The energy injected per unit mass $\vert\Delta\varepsilon\vert$ into the minihalo is given by Refs.~\cite{spitzer1958disruption, greenGoodwin2007mini}:
	\begin{equation}\label{eq:energy injected per unit mass (part 1)}
		\vert\Delta\varepsilon(r)\vert = \frac{\Delta E}{M_{\rm vir}} \frac{r^2}{\langle r^2 \rangle}
	\end{equation}
	where the net energy injected into the minihalo within the virial radius for a single encounter with a star of mass $m_{\ast}$, relative velocity $v_{\ast}$, and impact parameter $b$ is given by \cite{greenGoodwin2007mini}
	\begin{equation}
		\label{eq:deltaE_general}
		\Delta E = \begin{cases} 
			\dfrac{4 \alpha^{2}(c)}{3} \dfrac{G^{2} m_{\ast}^{2} M_{\rm vir} r^{2}_{\rm vir}}{v_{\ast}^2} \dfrac{1}{b^4} & \text{$(b>b_{\rm s})$} \\
			\dfrac{4 \alpha^{2}(c)}{3} \dfrac{G^{2} m_{\ast}^{2} M_{\rm vir} r^{2}_{\rm vir}}{v_{\ast}^2} \dfrac{1}{b_{\rm s}^4} & \text{$(b\leq b_{\rm s})$}
		\end{cases}
	\end{equation}
	where $b_{\rm s} = f_{\rm b}\, (2\alpha/3\beta)^{1/2} r_{\rm vir}$ is the transition radius and $f_{\rm b}$ is an order-unity correction factor introduced by S2023. From their simulations, S2023 finds that $f_{\rm b} = 6$.
	
	From Eqs.~(\ref{eq:langle r^2 rangle}), (\ref{eq:binding energy}) and (\ref{eq:energy injected per unit mass (part 1)}), it follows that:
	\begin{equation}\label{eq:energy injected per unit mass (part 2)}
		\vert\Delta\varepsilon(r)\vert = \Psi_0 \frac{\gamma}{\alpha^2} E_{\rm frac} \frac{r^2}{r_{\rm vir}^2}
	\end{equation}
	\begin{equation}\label{eq:normalized energy injected per unit mass (part 2) as a function of x}
		\vert\Delta\epsilon(x)\vert \equiv \frac{\vert\Delta\varepsilon(x)\vert}{\Psi_0}  = \frac{\gamma}{\alpha^2} E_{\rm frac} x^2
	\end{equation}
	where $\Psi_0 \equiv \frac{GM_{\rm vir}}{r_{\rm vir}}$, $E_{\rm frac} \equiv \frac{\Delta E}{E_{\rm b}}$, $G$ is Newton's constant, and $\vert\Delta\epsilon\vert$ is the normalized injected energy per unit mass into the minihalo.
	
	The relative potential $\Psi$ is defined as $\Psi(r) \equiv -\Phi(r)$, where $\Phi(r)$ is the Newtonian gravitational potential. For an untruncated NFW minihalo of concentration parameter $c$, it can be shown that (see Appendix \ref{app:psi for untruncated NFW})
	\begin{equation}\label{eq:Psi expression for NFW minihalo}
		\Psi(r) = \Psi_0\frac{1}{f_{\rm NFW}(c)} \frac{\ln\left(1+c\frac{r}{r_{\rm vir}}\right)}{\frac{r}{r_{\rm vir}}}
	\end{equation}
	\begin{equation}\label{eq:Psi expression for NFW minihalo as a function of x}
		\psi(x) \equiv \frac{\Psi(x)}{\Psi_0} = \frac{1}{f_{\rm NFW}(c)} \frac{\ln\left(1+cx\right)}{x}
	\end{equation}
	where $\psi$ is the normalized relative potential and
	\begin{equation}
		\label{eq:fNFW}
		f_{\rm NFW}(x) \equiv \ln(1+x) - \frac{x}{1+x}.  
	\end{equation}

	The expression for the mass loss in a minihalo due to a stellar encounter is (K2021):
	\begin{equation}\label{eq:mass loss NFW}
		\Delta M = 16\pi^2\int\limits_{r=0}^{r_{\rm vir}} \mathrm{d}r\ r^2 \int\limits_{\varepsilon=0}^{\min\left[\vert\Delta\varepsilon(r)\vert, \Psi(r)\right]} \mathrm{d}\varepsilon\sqrt{2(\Psi(r)-\varepsilon)}f(\varepsilon)
	\end{equation}
	where $\Delta M$ is the total mass loss within the virial radius of the minihalo. Also, $\varepsilon$ is a dark matter particle's specific relative (total) energy for a given $r$ and velocity. Additionally, $f(\varepsilon)$ is the phase space distribution function of dark matter particles in the minihalo (K2021).
	
	Converting Eq.~(\ref{eq:mass loss NFW}) to a dimensionless form, we compute the survival fraction (SF) of the minihalo as (see Appendix \ref{app:survival fraction expression for NFW minihalo})
	\begin{align}\label{eq:surival fraction expression dimensionless double integral}
		\text{SF} &\equiv 1 - \frac{\Delta M}{M_{\rm vir}} \nonumber\\
		&= 1 - \frac{4\pi c^3}{f_{\rm NFW}(c)}\int\limits_{x=0}^1\mathrm{d}x\ x^2\int\limits_{\epsilon=0}^{\min[\vert\Delta\epsilon(x)\vert, \psi(x)]} \mathrm{d}\epsilon\ \hat{f}(\epsilon)\sqrt{2(\psi(x)-\epsilon)}
	\end{align}
	where $\epsilon \equiv \frac{\varepsilon}{\Psi_0}$ is the normalized specific relative (total) energy and $\hat{f}(\epsilon) = \frac{\Psi_0^{3/2}}{\rho_{\rm s}} f(\varepsilon)$ (K2021) is the normalized phase space distribution function of the dark matter particles in the minihalo and it can be evaluated as follows (K2021): 
	\begin{equation}\label{eq:f_hat}
		\hat{f}(\epsilon) = \frac{1}{\sqrt{8}\pi^2} \int\limits_{\psi=0}^\epsilon \frac{1}{\sqrt{\epsilon - \psi}} \frac{\mathrm{d}^2\varrho}{\mathrm{d}\psi^2} \mathrm{d}\mathrm{\psi}
	\end{equation}
	where $\varrho \equiv \frac{\rho}{\rho_s}$ is the normalized density of the minihalo. Thus, Eq.~(\ref{eq:surival fraction expression dimensionless double integral}) can be rewritten as a triple integral:
	\begin{multline}\label{eq:surival fraction expression dimensionless triple integral}
		\text{SF} = 1 - \frac{4\pi c^3}{f_{\rm NFW}(c)}\int\limits_{x=0}^1 \int\limits_{\epsilon=0}^{\min[\vert\Delta\epsilon(x)\vert, \psi(x)]} \int\limits_{\psi^\prime=0}^\epsilon x^2 \frac{1}{\sqrt{8}\pi^2} \frac{1}{\sqrt{\epsilon - \psi^\prime}}\\ \frac{\mathrm{d}^2\varrho}{\mathrm{d}\psi^{\prime^2}}  \sqrt{2(\psi(x)-\epsilon)}\ \mathrm{d}\psi^\prime\ \mathrm{d}\epsilon\ \mathrm{d}x
	\end{multline}

\begin{figure}
		\includegraphics[width=\columnwidth]{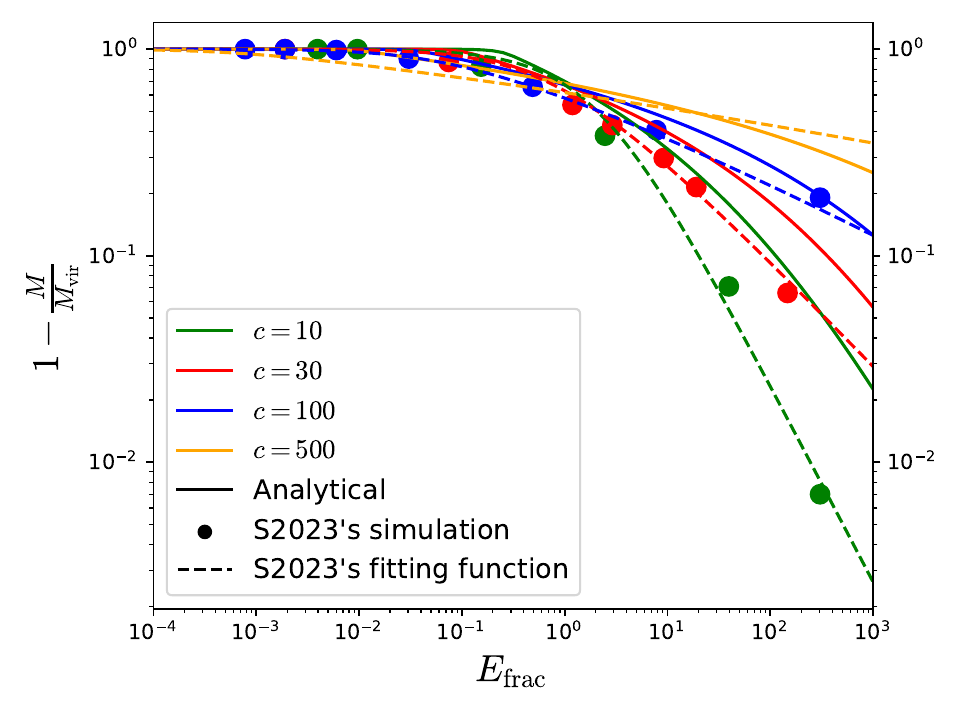}
		\caption{Survival fraction (SF) as a function of the normalized total injected energy $E_{\rm frac}$ into the minihalo as a result of stellar interaction, for concentration parameters $c=10,30,100,500$. The solid curves are the output of our implementation of K2021's analytical approach.   
			The dots are numerical simulation data from S2023. The dashed curves are the empirical fitting functions used by S2023.}
		\label{fig:SF vs Efrac for four values of concentration}
	\end{figure}

	Using Eq.~(\ref{eq:surival fraction expression dimensionless triple integral}), one can evaluate mass loss by computing the survival fraction of the minihalo for a particular value of $E_{\rm frac}$ and concentration parameter. We used the \textit{Derivative}() function from Python's SymPy library to evaluate $\frac{\mathrm{d}^2\varrho}{\mathrm{d}\psi^{\prime^2}}$ as a function of $x$ and the \textit{solve}() function from the SymPy library to invert the expression for $\psi^\prime$ so that we could express $x$ as a function of $\psi^\prime$ and eventually express $\frac{\mathrm{d}^2\varrho}{\mathrm{d}\psi^{\prime^2}}$ as a function of $\psi^\prime$. We further used the \textit{nquad}() function from the SciPy library to numerically evaluate the triple integral in Eq.~(\ref{eq:surival fraction expression dimensionless triple integral}). The survival fraction of the minihalo is then plotted against the normalized total injected energy, $E_{\rm frac}$, for a fixed concentration parameter, $c$. Fig.~\ref{fig:SF vs Efrac for four values of concentration} shows this plot for $c = 10,30,100,500$.
	From Fig.~\ref{fig:SF vs Efrac for four values of concentration}, it is clear that the analytical method described so far does not approximate the simulated data very well except for the $c=100$ case. To improve on the analytical method, we introduce 
	a sequential stripping model for the mass loss in the minihalo.
	
	\section{The sequential stripping model}
	\label{sec:The_sequential_stripping_model}
	One of the flaws with the expression for $\psi(x)$ as given by Eq.~(\ref{eq:Psi expression for NFW minihalo as a function of x}) is that it assumes that when one takes a dark matter particle from position $x$ to $\infty$, all matter in regions greater than normalized radius $x$ remains intact. 
	
	We introduce a model of dark matter particle ``unbinding," where we divide the minihalo into shells of infinitesimal thickness. During a stellar interaction, dark matter particles that will eventually be unbound will go to infinity. A dark matter particle in a particular shell that is going to infinity is not expected to feel a gravitational pull from a dark matter particle in an outer shell because shell expansion is taking place in a spherically symmetric manner. Thus, we model this as first starting with the minihalo's outermost shell (not necessarily within the virial radius) and taking it to infinity. Then we take the next innermost shell to infinity, and so on. When we take a dark matter particle from a normalized radius $x$ to infinity, we assume that no matter exists in the region of radius $>x$ because the matter in this region has already been taken to infinity. We call this approach the ``sequential stripping model".

\begin{figure}
		\includegraphics[width=\columnwidth]{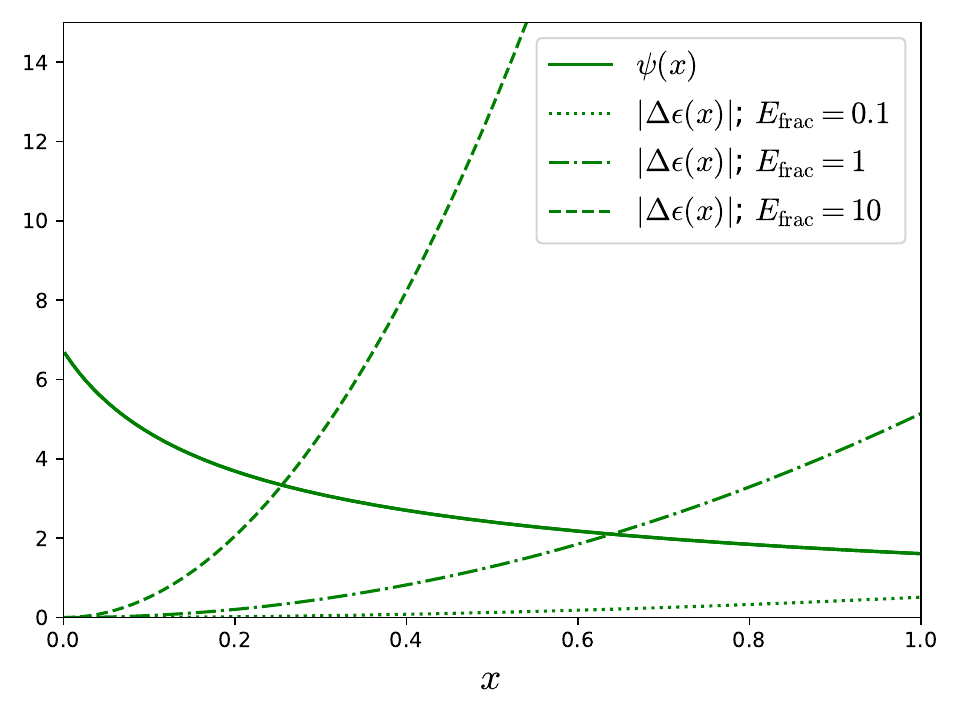}
		\caption{The normalized injected energy per unit mass $\vert\Delta\epsilon\vert$ in a minihalo due to a stellar interaction is plotted as a function of the normalized radius $x$. In doing so, the total injected energy $E_{\rm frac}$ is set to 0.1, 1, 10. 
			The normalized relative potential $\psi$ is also plotted as a function of $x$. The $\psi(x)$ curve does not vary with $E_{\rm frac}$. In all cases for this figure, the concentration parameter $c=10$.  The value of $x$ at which $\vert\Delta\epsilon(x)\vert$ and $\psi(x)$ curves intersect is called the normalized crossover radius $x^*$.}
		\label{fig:psi crossing over deltaEpsilon}
	\end{figure}

	In Eq.~(\ref{eq:surival fraction expression dimensionless triple integral}), the upper limit of the $\epsilon$ is a \textit{minimum} operation on two functions, $\psi(x)$ and $\vert\Delta\epsilon(x)\vert$. Setting $c=10$ for now, Fig.~\ref{fig:psi crossing over deltaEpsilon} shows $\psi(x)$ (according to Eq.~(\ref{eq:Psi expression for NFW minihalo as a function of x})) which is a decreasing function of $x$ and $\vert\Delta\epsilon(x)\vert$ (according to Eq.~(\ref{eq:normalized energy injected per unit mass (part 2) as a function of x})) which is a quadratically increasing function on $x$. The two curves intersect within or beyond the virial radius, and the normalized radius of the intersection is called the normalized crossover radius $x^*$.
	To make Eq.~(\ref{eq:surival fraction expression dimensionless triple integral}) more tractable, let's utilize the normalized crossover radius, $x^*$, which is mathematically defined as that value of $x$ such that:
	\begin{equation}\label{eq:cross over radius condition}
		\vert\Delta\epsilon(x^*)\vert = \psi(x^*)
	\end{equation}
	
	As shown in Appendix~(\ref{app:Evaluating the survival fraction using the sequential stripping model}) the survival fraction with the sequential stripping model can be computed using
	\begin{equation}\label{eq:SF expression in terms of mass fraction x < x^* and I_A copy}
		\text{SF} = \text{mass fraction}_{x < \min[x^*,1]} - \text{prefactor}\times I_{\rm A}
	\end{equation}
	where
	\begin{equation}\label{eq:expression for mass fraction x < x^*, in terms of x copy}
		\text{mass fraction}_{x < \min[x^*,1]} = \frac{c^2}{f_{\rm NFW}(c)} \int\limits_{x=0}^{\min[x^*, 1]} \frac{x}{\left(1 + cx\right)^2} \mathrm{d}x,
	\end{equation}
	\begin{equation}\label{eq:prefactor copy}
		\text{prefactor} \equiv \frac{4\pi c^3}{f_{\rm NFW}(c)}
	\end{equation}
	and 
	\begin{multline}\label{eq:I1 updated to sequential stripping model copy}
		I_{\rm A} \equiv \int\limits_{x=0}^{\min[x^*, 1]} \int\limits_{\epsilon=0}^{\vert\Delta\epsilon(x)\vert} \  \int\limits_{\psi_{\rm B}^\prime=0}^\epsilon x^2\frac{1}{\sqrt{8}\pi^2}\ \frac{1}{\sqrt{\epsilon - \psi_{\rm B}^\prime}} \frac{\mathrm{d}^2\varrho}{\mathrm{d}\psi_{\rm B}^{\prime^2}}\left(x^\prime\left(\psi_{\rm B}^\prime\right)\right)\\ \sqrt{2\left(\psi_{\rm A}(x)-\epsilon\right)} \ \mathrm{d}\psi_{\rm B}^\prime\ \mathrm{d}\epsilon\ \mathrm{d}x.
	\end{multline}
	The relative potentials in the above equation are given by 
	\begin{align}\label{eq:psi1(x) copy}
		\psi_{\rm A}(x) = \frac{1}{f_{\rm NFW}(c)} \left[ \frac{\ln(1+cx)}{x} - \frac{c}{1+cx^*} \right] && ,x<x^*
	\end{align}
	and
	\begin{align}\label{eq:psi2(x) copy}
		\psi_{\rm B}(x) = \frac{1}{f_{\rm NFW}(c)} \left[ \frac{\ln(1+cx)}{x} - \frac{c}{1+cx} \right] && ,x>x^*.
	\end{align}
	
\begin{figure*}
		\includegraphics[width=\textwidth]{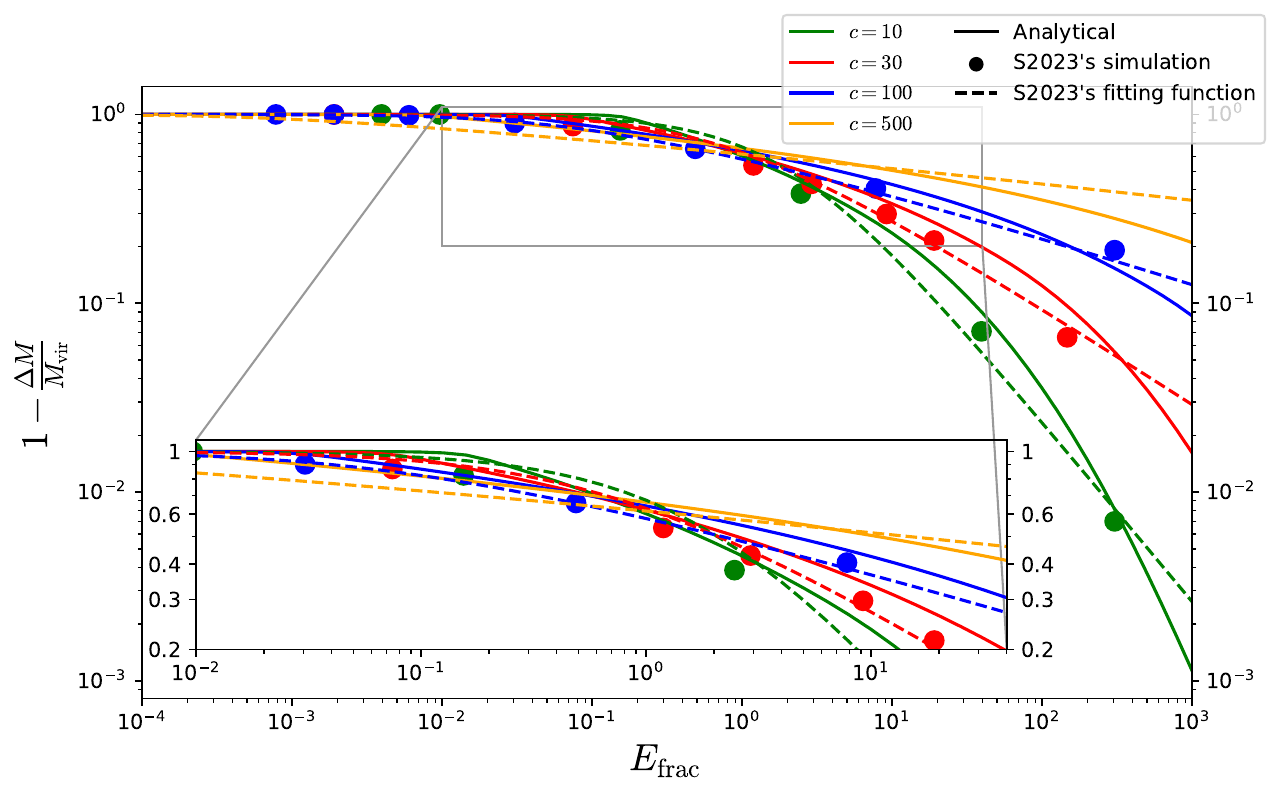}
		\caption{ The same as Fig.~\ref{fig:SF vs Efrac for four values of concentration} except that
			the solid curves are the output of our analytical approach using the sequential stripping model of mass loss in the minihalo.
		}
		\label{fig:SF vs Efrac - sequential stripping model}
	\end{figure*}

	Fig.~\ref{fig:SF vs Efrac - sequential stripping model} shows how the survival fraction (solid curves) varies with the normalized injected energy $E_{\rm frac}$. The sequential stripping model gives a better fit to the simulation data points when compared to the approach used in Fig.~\ref{fig:SF vs Efrac for four values of concentration}.

	\subsection{The need to include relaxation}\label{sec:need for relaxation}
	The S2023 simulation data shown in our  Fig.~\ref{fig:SF vs Efrac - sequential stripping model}, is computed after the remnant minihalo has undergone full relaxation following a stellar encounter. However, we have not accounted for this relaxation process for our analytical curves. But by juxtaposing our analytical curves with S2023's numerical data points, it is equivalent to assuming that, in our case, the remnant minihalo relaxes to the same initial NFW profile. This is a good approximation for small $E_{\rm frac}$ values. For example, in the limit $E_{\text{frac}} \rightarrow 0$, there is no actual stellar encounter, and the resulting (relaxed) minihalo is the same NFW minihalo we started with. Thus, for small finite $E_{\rm frac}$ values, the approximation that the remnant minihalo relaxes to the same NFW profile is a good one. Thus, we see that our analytical curves have a good match to the numerical data at low values of $E_{\rm frac}$. However, S2023 finds that, in general, the remnant minihalo relaxes to a broken power profile,
	\begin{equation}
		\rho(r)=\frac{\rho_s}{\frac{r}{r_s}\left(1+\frac{r}{r_s}\right)^k},
	\end{equation}
	which has an outer \cg{logarithmic} slope of \cg{$-(k+1)$} for large values of $r$ and a \cg{logarithmic} slope of \cg{$-1$} for small values of $r$ \footnote{This is in contrast to Ref.~\cite{Delos:2019tsl} who finds that the profiles relax to a non-broken power law formula. However, Ref.~\cite{Delos:2019tsl} only fits numerical simulations up to $r\approx 2 r_s$ while S2023 fits them to $r\approx 20 r_s$.}. For our purposes, we assume that $k=3$, and this profile is known as the Hernquist density profile \cite{Hernquist:1990be}. Since we have argued that for small $E_{\rm frac}$, the remnant minihalo relaxes to an NFW profile, it is only for larger $E_{\rm frac}$ values that the remnant minihalo relaxes to a Hernquist profile. Since our analytical curves in Fig.~\ref{fig:SF vs Efrac - sequential stripping model} don't account for this, there arises a discrepancy which becomes starker at higher values of $E_{\rm frac}$. This discrepancy leads to errors, which result in our analytical curves overshooting the numerical data points at larger $E_{\rm frac}$ values. We need to include the relaxation process in our calculations to account for this discrepancy.
	
	\subsection{The Hernquist model}
	In this section, we assume that after disruption, the minihalo relaxes to the Hernquist density profile, which is described by
	\begin{equation}\label{eq:Hernquist profile in terms of r}
		\rho(r) = \frac{\rho_{\rm s}}{\frac{r}{r_{\rm s}} \left( 1 + \frac{r}{r_{\rm s}} \right)^3}
	\end{equation}
	\begin{equation}\label{eq:Hernquist profile in terms of x}
		\rho(x) = \frac{\rho_{\rm s}}{cx \left( 1 + cx \right)^3}
	\end{equation}
	
	Firstly, using the same sequential stripping model we used for the NFW minihalo, it can be shown that the normalized relative potential of an untruncated Hernquist minihalo is (see Appendix~\ref{app:psi for Hernquist minihalo})
	\begin{align}\label{eq:psI_A(x) for Herquist Profile}
		\psi_{\rm A}(x) = (1+c)^2 \biggr[& \frac{x^* - x}{(1+cx^*)(1+cx)} \nonumber\\
		&+ \frac{x^*}{(1+cx^*)^2} \biggr]&& ,  x<x^*
	\end{align}
	\begin{align}\label{eq:psI_B(x) for Herquist Profile}
		\psi_{\rm B}(x) = (1+c)^2 \frac{x}{(1+cx)^2}&& ,  x>x^*
	\end{align}

	To find $x^*$, we use the condition given in Eq.~(\ref{eq:cross over radius condition}) where $\vert\Delta\epsilon(x)\vert$ is given by Eq.~(\ref{eq:normalized energy injected per unit mass (part 2) as a function of x}). The $\alpha^2$ and $\gamma$ for a Hernquist minihalo are given by (see Appendix~\ref{app:alpha squared and beta for Herquist minihalo})
	\begin{equation}\label{eq:alpha squared for Herquist minihalo}
		\alpha^2 = \frac{c(6 + 9c + 2c^2) - 6(1 + c)^2\ln(1 + c)}{c^4}
	\end{equation}
	\begin{equation}\label{eq:beta for Hernquist minihalo}
		\gamma = \frac{4+c}{6}
	\end{equation}

	\subsection{Density profile of the first-generation minihalo resulting from stellar interaction with an NFW minihalo}\label{sec:density profile of first-generation minihalo after single encounter}\label{sec:density profile of first-generation minihalo}
	We would now like to plot the density profile of the first-generation minihalo resulting from a stellar interaction of an NFW minihalo. The first-generation minihalo is assumed to have a broken power law profile (S2023). We start by specifying the scale parameters of an unperturbed NFW minihalo. Using these, we look to compute the scale parameters of the first-generation broken power law profile minihalo. We use the subscript $s$ to denote the unperturbed NFW minihalo and subscript $1$ to denote the resulting first-generation minihalo.
	
	We first note that the NFW minihalo in the region $x>x_{\rm s}^*$ is completely disrupted. In addition, there is a partial mass loss in the region $x<x_{\rm s}^*$ \cg{(see Appendix~\ref{app:Evaluating the survival fraction using the sequential stripping model})}. One must note that $x_{\rm s}^*$ can be greater than 1. We first compute the total surviving mass of the NFW minihalo just after the stellar interaction. We then allow the remnant minihalo to fully relax to a broken power law profile of the form
	\begin{equation}\label{eq:general broken powerlaw profile}
		\rho(r)=\frac{\rho_1}{\frac{r}{r_1}\left(1+\frac{r}{r_1}\right)^k}    
	\end{equation}
	where we are leaving $k$ unspecified for the moment, and later we will set it to 3.
	We use the fact that the total surviving mass of the NFW minihalo after perturbation is equal to the total mass of the fully relaxed broken power law first-generation minihalo.
	i.e.,
	\begin{equation}\label{eq:mass condition for transition from NFW minihalo to first-generation minihalo}
		M_{\text {enc,s}}\left(x_{\rm s}^*\right)-\Delta M_{x_{\rm s}=0 \rightarrow x_{\rm s}^*}=\lim _{x_1 \rightarrow \infty} M_{\rm enc, 1}(x_1)
	\end{equation}
	where $M_{\text {enc,s}}\left(x_{\rm s}^*\right)$ is the mass of the NFW minihalo enclosed within the normalized crossover radius $x_{\rm s}^*$ and $\Delta M_{x_{\rm s}=0 \rightarrow x_{\rm s}^*}$ is the mass of the NFW minihalo lost within the normalized crossover radius $x_{\rm s}^*$. Also, $\lim\limits_{x_1 \rightarrow \infty} M_{\rm enc, 1}(x_1)$ is the total mass of the first-generation broken power law minihalo. The $x_{\rm s}$ and $x_1$ are ``local variables'' of the NFW minihalo and the first-generation minihalo, respectively. They are defined as
	\begin{equation}\label{eq:x_s defintion}
		x_{\rm s}\equiv\frac{r}{r_{\text {vir,s}}}
	\end{equation}
	\begin{equation}\label{eq:x_1 defintion}
		x_1\equiv\frac{r}{r_{\text {vir,1}}}
	\end{equation}
	where $r_{\text {vir,s}}$ and $r_{\text {vir,1}}$ are the virial radii of the unperturbed NFW minihalo and first-generation minihalo, respectively.
	
	We consider the disrupted halo to be  a broken power law of the form 
	\begin{equation}
		\label{eq:broken power law}
		\rho_{k=2+\Delta}(r)=\frac{\rho_1}{\frac{r}{r_1}\left(1+\frac{r}{r_1}\right)^{2+\Delta}}
	\end{equation}
	for a parameter $\Delta > 0$.

	We now look at Fig. 6 of S2023 and make the reasonable assumption that at small radii, the unperturbed NFW minihalo density profile and the first-generation Hernquist/broken power-law density profile are indistinguishable from each other. We write this as
	\begin{align}\label{eq:small radius condition}
		\lim _{r \rightarrow 0} \rho_{\rm NFW}(r)&=\lim _{r \rightarrow 0} \rho_{k=2+\Delta}(r) \nonumber\\
		\lim _{r \rightarrow 0} \frac{\rho_{\rm s}}{\frac{r}{r_{\rm s}}\left(1+\frac{r}{r_{\rm s}}\right)^2}&=\lim _{r \rightarrow 0} \frac{\rho_1}{\frac{r}{r_1}\left(1+\frac{r}{r_1}\right)^{2+\Delta}} \nonumber\\
		\frac{\rho_{\rm s}}{\frac{r}{r_{\rm s}}}&=\frac{\rho_1}{\frac{r}{r_1}} \nonumber\\
		\Rightarrow \rho_{\rm s} r_{\rm s}&=\rho_1 r_1
	\end{align}
	
	As shown in Appendix~\ref{app:Computing expressions for the disrupted minihalo's parameters}
	we can  compute $r_1$ in terms of $r_{\rm s}$ as follows
	\begin{align}\label{eq:r_1 final expression - broken power law k=3.2 copy}
		r_1
		&=r_{\rm s} \times R_{\rm s}
	\end{align}
	where
	\begin{equation}\label{eq:r_s defintion - broken power law k=3.2 copy}
		R_{\rm s} \equiv \sqrt{(\Delta + \Delta^2)\left[f_{\rm NFW}(c_{\rm s} x_{\rm s}^*)-4 \pi c_{\rm s}^3 I_{\rm s}\right]}
	\end{equation}
	and 
	\begin{equation}\label{eq:c_s definition copy}
		c_{\rm s} \equiv \frac{r_{\rm vir,s}}{r_{\rm s}}.
	\end{equation}
	It is also shown in the same appendix that 
	\begin{equation}\label{eq:rho_1 final condition copy}
		\rho_1 = \frac{\rho_{\rm s}}{R_{\rm s}}
	\end{equation}
	Thus, if the scale parameters $\rho_{\rm s}$, $r_{\rm s}$ of the NFW profile are given, Eqs.~(\ref{eq:r_1 final expression - broken power law k=3.2 copy}) and (\ref{eq:rho_1 final condition copy}) give the scale parameters $\rho_1$, $r_1$ of the resulting first-generation broken power law minihalo.

	Using N-body simulations, S2023 finds that when an NFW minihalo participates in a stellar interaction with impact parameter $b=2 \times 10^{-5} \mathrm{kpc}$, the resulting first-generation minihalo will have a broken power law profile with $k=3.2$ or $\Delta = 1.2$. In such a case,
	\begin{equation}
		R_{\rm s} = \sqrt{\frac{66}{25}\left[f_{\rm NFW}(c_{\rm s} x_{\rm s}^*)-4 \pi c_{\rm s}^3 I_{\rm s}\right]}
	\end{equation}
	On the other hand, according to the  N-body simulations of S2023, if the impact parameter is $b=5 \times 10^{-5} \mathrm{kpc}$, the resulting first-generation minihalo will have a broken power law profile with $k=3.3$ or $\Delta = 1.3$. Then,
	\begin{equation}
		R_{\rm s} = \sqrt{\frac{299}{100}\left[f_{\rm NFW}(c_{\rm s} x_{\rm s}^*)-4 \pi c_{\rm s}^3 I_{\rm s}\right]}
	\end{equation}
	Finally, assuming for simplicity, that the first-generation minihalo had a Hernquist profile ($k = 3$ or $\Delta = 1$),
	\begin{equation}\label{eq:r_s definition}
		R_{\rm s} = \sqrt{2f_{\rm NFW}(c_{\rm s} x_{\rm s}^*) - 8 \pi c_{\rm s}^3 I_{\rm s}}
	\end{equation}
	Moving on, instead of specifying the unperturbed NFW minihalo by its two scale parameters, we would like to specify it by two other quantities: its concentration parameter and its virial mass. So, in order to compute the scale parameters of the first-generation minihalo, we need to first compute the scale parameters of the NFW minihalo from its concentration and virial mass.
	
	We first use the definition of the virial radius of the NFW minihalo. The virial radius is that radius at which the average density $\bar{\rho}_{\rm vir}$ enclosed within the virial radius is 200 times the critical density $\rho_{\rm crit}$ of the universe. Therefore,
	\begin{align}\label{eq:Virial radius condition}
		\frac{\text{Mass within virial radius}}{\text{Volume within virial radius}} &= 200\rho_{\rm crit} \nonumber\\\
		\frac{\int\limits_{r=0}^{r_{\rm vir, s}} \rho_{\rm NFW}(r) \times 4 \pi r^2 d r}{\frac{4 \pi}{3} r_{\rm vir, s}^3}&=200 \rho_{\text {crit }}
	\end{align}
	Let
	\begin{equation}\label{eq:x_s definition}
		x_s \equiv \frac{r}{r_{\text {vir,s}}}\ \Rightarrow\ r=x_{\rm s} r_{\text {vir,s }}
	\end{equation}
	Substituting Eq.~(\ref{eq:x_s definition}) in Eq.~(\ref{eq:Virial radius condition})
	\begin{equation}\label{eq:Virial radius condition, intermediate 2}
		\int\limits_{x_{\rm s}=0}^1 \rho_{\rm NFW}\left(x_{\rm s}\right) x_{\rm s}^2 \mathrm{d} x_{\rm s}=\frac{200}{3} \rho_{\text {crit}}
	\end{equation}
	Substituting Eq.~(\ref{eqn:NFW as a function of x}) in Eq.~(\ref{eq:Virial radius condition, intermediate 2}) and performing the integration with respect to $x_{\rm s}$ in the L.H.S. of Eq.~(\ref{eq:Virial radius condition, intermediate 2}), we get
	\begin{equation*}
		\frac{\rho_{\rm s}}{c^3}\left[\ln (1+c)-\frac{c}{1+c}\right]=\frac{200}{3} \rho_{\rm crit}
	\end{equation*}
	\begin{equation}\label{eq:rho_s in terms of concentration for NFW profile}
		\Rightarrow \rho_{\rm s}=\frac{200}{3} \frac{c^3}{f_{\rm NFW}(c)} \rho_{\rm crit} 
	\end{equation}
	Cosmological observations \citep{Planck:2018vyg} fix the value of $\rho_{\rm crit}$ as
	\begin{equation}\label{eq:rho_crit value}
		\rho_{\text {crit}}=1.3483 \times10^{-7} \mathrm{M}_\odot / \mathrm{pc}^3
	\end{equation}
	Thus, specifying the concentration of the NFW minihalo fixes its scale density through Eq.~(\ref{eq:rho_s in terms of concentration for NFW profile}).
	
	Next, the average density within the virial radius of the NFW minihalo is expressed in terms of its virial mass and viral radius as follows
	\begin{equation}\label{eq:average virial density definition}
		\bar{\rho}_{\text {vir}}=\frac{M_{\text {vir,s}}}{\frac{4 \pi}{3} r_{\text {vir,s}}^3}    
	\end{equation}

	Substituting Eq.~(\ref{eq:c_s definition copy}) in Eq.~(\ref{eq:average virial density definition}) and solving for $r_{\rm s}$, we get
	\begin{equation}\label{eq:r_s in terms of of c_s and M_vir,s - pre-final expression}
		r_{\rm s}=\frac{1}{c_{\rm s}}\left(\frac{3 M_{\text {vir,s}}}{4 \pi \bar{\rho}_{\text {vir}}}\right)^{1 / 3}
	\end{equation}
	Noting that by definition
	\begin{equation}
		\bar{\rho}_{\rm vir}=200 \rho_{\text {crit}}    
	\end{equation}

	Therefore, Eq.~(\ref{eq:r_s in terms of of c_s and M_vir,s - pre-final expression}) becomes
	\begin{equation}\label{eq:r_s in terms of of c_s and M_vir,s - final expression}
		r_{\rm s}=\frac{1}{c_{\rm s}}\left(\frac{3 M_{\rm vir,s}}{800 \pi \rho_{\text {crit}}}\right)^{1 / 3} 
	\end{equation}
	Thus, given the concentration and virial mass of the NFW minihalo, its scale parameters can be found using Eqs.~(\ref{eq:rho_s in terms of concentration for NFW profile}) and (\ref{eq:r_s in terms of of c_s and M_vir,s - final expression}).
	\begin{figure*}[htp]
		
		\subfloat{
			\includegraphics[clip,width=\columnwidth]{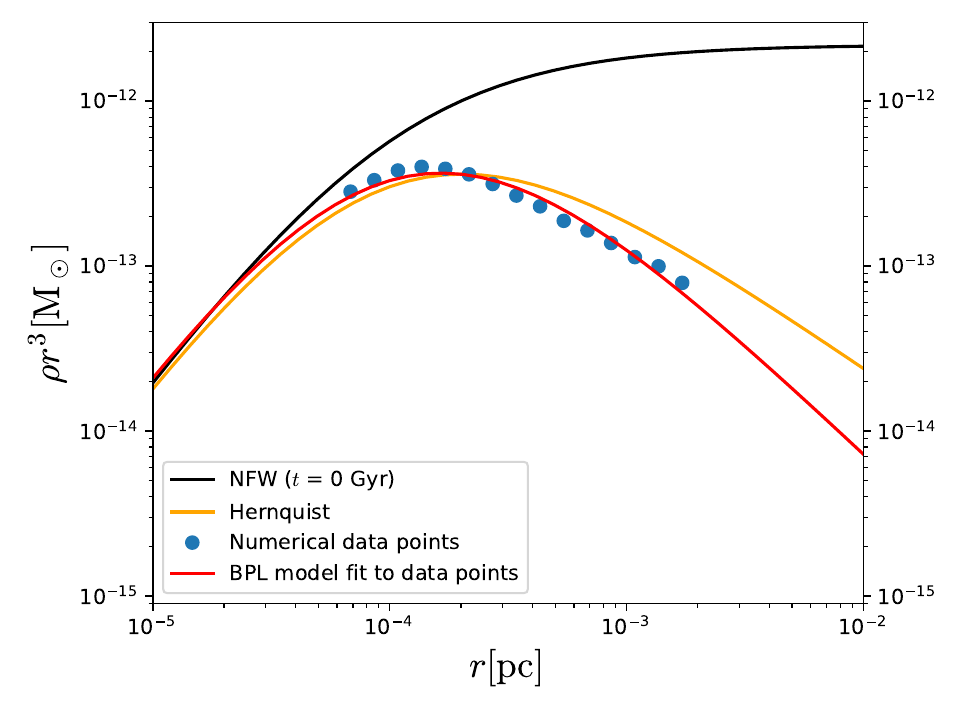}
		}
		\subfloat{
			\includegraphics[clip,width=\columnwidth]{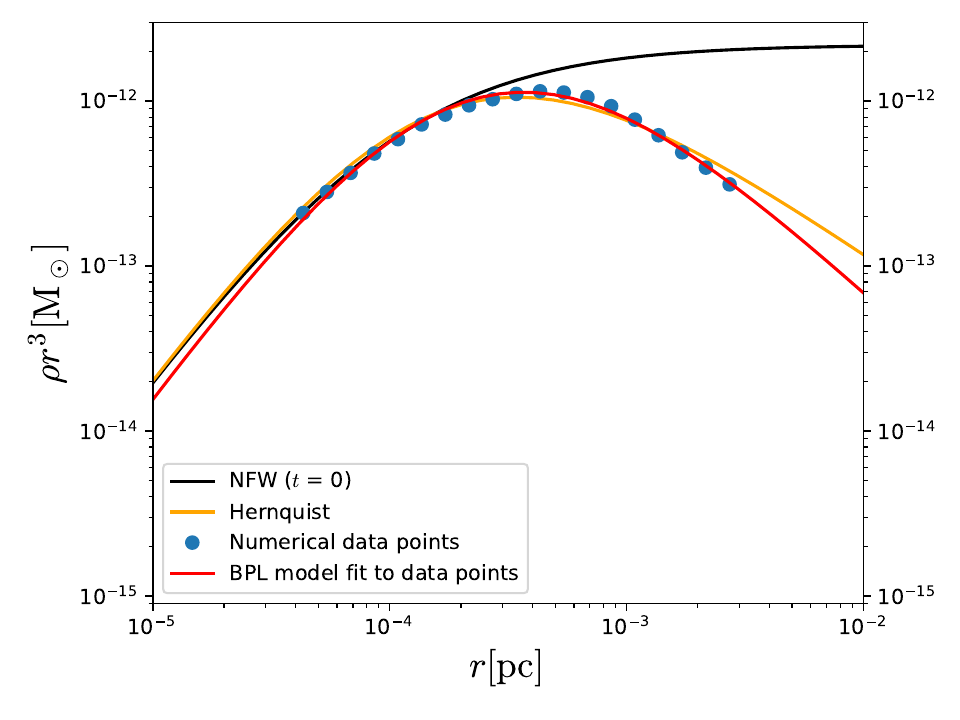}
		}
		\caption{The left and right panels show the density profile of an NFW minihalo (black curve) which has a stellar encounter with impact parameters $b = 2\times10^{-5}\rm kpc$ and $b = 5\times10^{-5}\rm kpc$ respectively. The NFW minihalo has an initial concentration \cg{$c_{\rm s}=100$} and virial mass $M_{\rm vir} = 10^{-10}\rm M_\odot$. The orange curve shows the case where the remnant minihalo has relaxed to a Hernquist density profile, which is a broken power law \cg{(BPL)} profile with $k=3$. The teal dots are numerical data points of the resulting density profile stabilized at $t = 2.5$ Gyr post-stellar interaction taken from S2023. We performed a curve fit of these data points to a broken power law profile and obtained its three parameters. The broken power law profile is shown as the red curves.}
		\label{fig:density profile of first-generation minihalo}
	\end{figure*}
	Given the concentration and virial mass of the unperturbed NFW minihalo, we are now able to plot the density profile of the first-generation minihalo. We use $c_{\rm s}=100$ and $M_{\text {vir,s}}=10^{-10} \rm M_\odot$. The left (right) panel of Fig.~\ref{fig:density profile of first-generation minihalo} shows the case where the impact parameter $b=2 \times 10^{-5} \mathrm{kpc}$ ($b=5 \times 10^{-5} \mathrm{kpc}$). The black curve shows the density profile of the unperturbed NFW minihalo while the orange curve shows the case where the fully relaxed first-generation minihalo is assumed to have a Hernquist profile $(k=3)$. For both the impact parameter cases, S2023 numerically shows how the density profile of the resulting minihalo changes and stabilizes over time. We chose the data points on that portion of the density profile curve at $t$ = 2.5 Gyr where the profile had stabilized and used these datapoints to find the model parameters of the resulting broken power law of the form given in Eq.~(\ref{eq:general broken powerlaw profile}). These data points are shown in teal in both panels of Fig.~\ref{fig:density profile of first-generation minihalo}. We did a least squares fit of the data points and obtained the optimal parameters $r_1$, $\rho_1$ and $k$. These parameters are given in Table~\ref{tab:BPL model parameters and survival fractions}. The resulting broken power law profiles are plotted as the red curves in Fig.~\ref{fig:density profile of first-generation minihalo}. We notice that the broken power law profile merges with the NFW profile at small radii. This further reinforces our assumption that the parent and relaxed child AMCs are indistinguishable at small radii during a stellar interaction. Next, we calculated the virial radius $r_{\rm vir,s}$ of the unperturbed NFW minihalo using Eq.~(\ref{eq:average virial density definition}). We found that $r_{\rm vir,s} = 9.6\times10^{-3}$pc. We then calculated the masses of the Hernquist and broken power law profiles inside this virial radius $r_{\rm vir,s}$. The ratio of the above mass to the virial mass of the NFW profile then gives us the survival fraction of the Hernquist or broken power-law profiles. Alternatively, we also calculate the value of normalized injected energy $E_{\rm frac}$ using the impact parameter \cg{with Eq.~(\ref{eq:Efrac expression in terms of encounter parameters}) when $b>b_{\rm s}$. When $b\leq b_{\rm s}$ one can use Eq.~(\ref{eq:Efrac expression in terms of encounter parameters}) with $b$ replaced by $b_{\rm s}$.} Knowing $E_{\rm frac}$, we calculate the survival fraction according to S2023's empirical response function. These survival fractions are presented in Table~\ref{tab:BPL model parameters and survival fractions}. As can be seen, there is about a 0.03 difference in the SF value between the broken power law case and the SF from S2023's empirical response function. This is indicative of the level of systematic error in determining the SF. As can be seen, the difference between the SF from the Hernquist and the broken power law profiles is around 0.01, which indicates that within the level of systematic error, the Hernquist profile could be used instead of the broken power law. 
	
	\begin{table}
		\begin{tabular}{|l|c|c|}
			\hline
			b &	$2\times10^{-5}$kpc & $5\times10^{-5}$kpc \\
			\hline
			$r_1$ & $1.14\times10^{-4}$pc & $2.47\times10^{-4}$pc \\
			\hline
			$\rho_1$ & $2.44$M$_\odot$/pc$^3$ & $0.716$M$_\odot$/pc$^3$\\
			\hline
			$k$  & 3.38 & 3.34\\
			\hline
			SF from Hernquist & 0.15 & 0.433 \\
			\hline
			SF from BPL & 0.137 & 0.423 \\
			\hline
			SF from S2023 response function & 0.172 & 0.393 \\
			\hline
		\end{tabular}
		\caption{
			\label{tab:BPL model parameters and survival fractions}
			An NFW profile minihalo of concentration $c_{\rm s}$ and virial mass $M_{\rm vir,s}$ undergoes a stellar interaction with two different impact parameters $b=2 \times 10^{-5} \mathrm{kpc}$ and $b=5 \times 10^{-5} \mathrm{kpc}$. In each case, we fit the numerical data points in Fig.~\ref{fig:density profile of first-generation minihalo} to obtain the parameters of the resulting broken power law (BPL) profile. The scale radius $r_1$, scale density $\rho_1$, and power parameter $k$ of the BPL profile are presented above. Moreover, the survival fractions using the Hernquist and BPL profiles in Fig.~\ref{fig:density profile of first-generation minihalo} as well as using S2023's empirical response function are presented.
		}
	\end{table}

	\subsection{Incorporating relaxation to a Hernquist profile}
	We now try to improve upon Fig.~\ref{fig:SF vs Efrac - sequential stripping model}, this time incorporating the relaxation process to a Hernquist profile. We assume we are given the concentration $c_{\rm s}$ of the unperturbed NFW minihalo. Then, Eq.~(\ref{eq:rho_s in terms of concentration for NFW profile}) gives us the scale density $\rho_{\rm s}$ of the NFW minihalo. Having calculated $\rho_{\rm s}$, Eq.~(\ref{eq:rho_1 final condition copy}) gives us the scale density $\rho_1$ of the relaxed first-generation Hernquist minihalo, where \cg{$R_{\rm s}$} is given by Eq.~(\ref{eq:r_s definition})
	
	Our next task is to compute the concentration $c_1$ of the first-generation Hernquist minihalo and get an equation analogous to Eq.~(\ref{eq:rho_s in terms of concentration for NFW profile}) but for the Hernquist profile. We start with the definition of the virial radius as before. This then leads us to an equation similar to Eq.~(\ref{eq:Virial radius condition, intermediate 2}) but for the first-generation Hernquist minihalo, finally resulting in an equation relating the concentration $c_1$ and scale density $\rho_1$ as follows (see Appendix~\ref{app:computing the concentration of Hernquist minihalo, given its scale density}): 
	
	\begin{equation}\label{eq:relating concentration and scale radius for Hernquist profile}
		\frac{1}{2 c_1\left(1+c_1\right)^2}=\frac{200}{3}\frac{\rho_{\text {crit}}}{\rho_1}
	\end{equation}
	The survival fraction after relaxation is given by (see Appendix~\ref{app:Computing the expression for survival fraction with relaxation})
	
	\begin{align}\label{eq:survival fraction of NFW minihalo incorporating relaxation}
		\text {SF} &\equiv \frac{M_{\text {enc,1}}\left(x_1^{r_{\rm vir,s}}\right)}{M_{\text {vir,s}}} \\
		& =\frac{1}{2} R_{\rm s}^2 \frac{f_{\rm Hern}(c_1 x_1^{r_{\text{vir,s}}})}{f_{\rm NFW}(c_{\rm s})} 
	\end{align}
	where
	\begin{equation}
		f_{\rm Hern}(x) = \frac{x^2}{(1 + x)^2}
	\end{equation}
	\begin{equation}
		x_1^{r_{\rm vir,s}} \equiv \frac{r_{\rm vir,s}}{r_{\rm vir,1}}
	\end{equation}
	is the virial radius of the unperturbed NFW minihalo expressed in terms of the ``local" normalized radial distance variable of the first-generation Hernquist minihalo. $M_{\rm enc,1}$ is the mass enclosed for a Hernquist profile.
	$x_1^{r_{\text{vir,s}}}$ can be written as (see Appendix~\ref{app:Computing the expression for survival fraction with relaxation})
	
	\begin{equation}
		x_1^{r_{\text{vir,s}}} = \frac{c_{\rm s}}{c_1} \frac{1}{R_{\rm s}}
	\end{equation}
	\begin{figure*}
		\includegraphics[width=\textwidth]{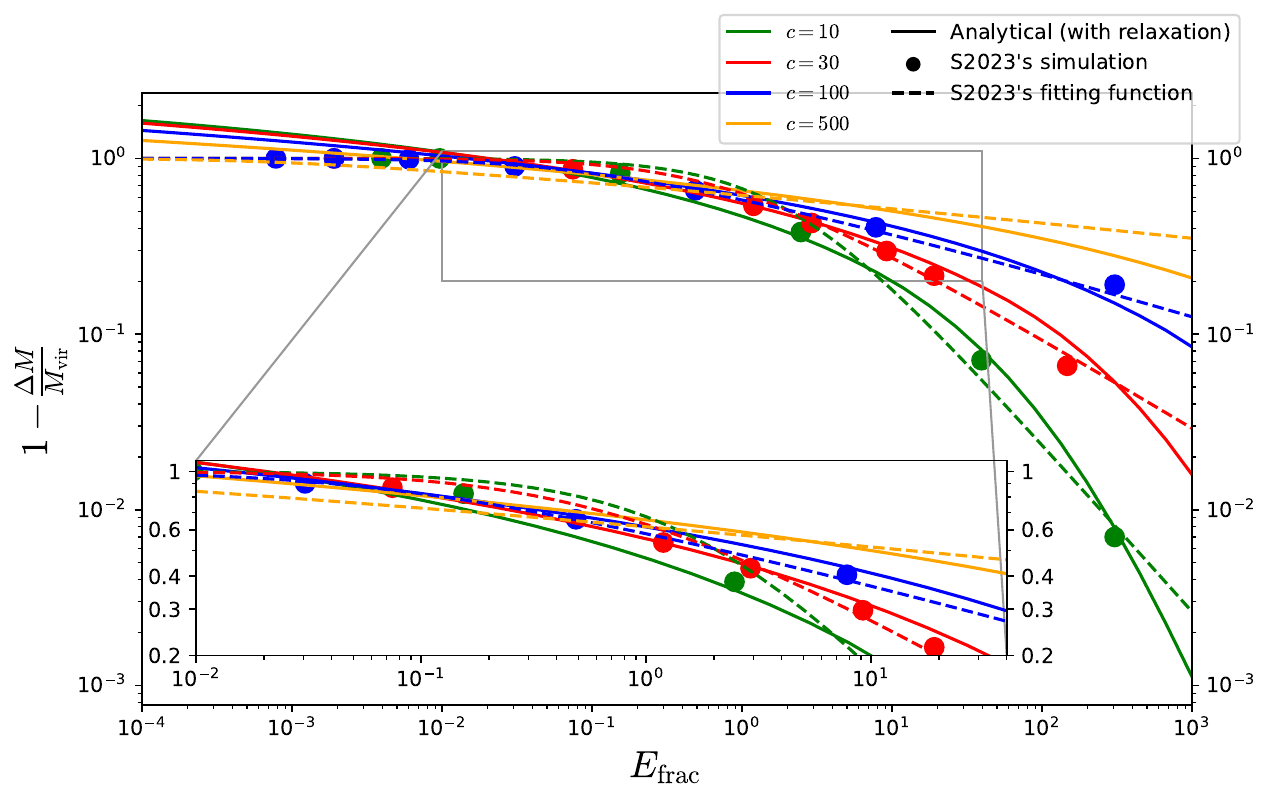}
		\caption{
			The same as Fig.~\ref{fig:SF vs Efrac for four values of concentration} except that the solid curves are the output of our analytical approach using the sequential stripping model of mass loss in the minihalo and taking into account relaxation of the remnant minihalo to a Hernquist profile.
		}
		\label{fig:SF vs Efrac - including relaxation}
	\end{figure*}
	Using this procedure, the survival fraction can be computed against $E_{\rm frac}$. Fig.~\ref{fig:SF vs Efrac - including relaxation} shows the results. The solid lines are our analytical curves. The circular dots are numerical data points from S2023. The dashed curves are S2023's curve fits to the numerical data. Our analytical curves closely match the numerical data points for larger values of $E_{\rm frac}$. This is because it is a reasonable assumption that the remnant minihalo relaxes to a Hernquist profile in this regime. However, our analytical curves overshoot the numerical data for smaller values of $E_{\rm frac}$. This is because, as discussed in Section~\ref{sec:need for relaxation}, for smaller $E_{\rm frac}$ values, it is a better approximation to assume that the remnant minihalo relaxes to the same NFW profile, not the Hernquist profile.
	
	\subsection{The switching procedure}
	Fig.~\ref{fig:SF vs Efrac - sequential stripping model} shows that the sequential striping model without considering relaxation produces a good match to the numerical data at low values of $E_{\rm frac}$ but overshoots the data at high $E_{\rm frac}$ values. On the other hand, Fig.~\ref{fig:SF vs Efrac - including relaxation} shows that the sequential stripping model incorporating relaxation to a Hernquist profile overshoots the numerical data at low values of $E_{\rm frac}$ but is a good match at high $E_{\rm frac}$ values. We have previously discussed the physical intuitions for both these models. We now introduce a switching procedure where, for any given $E_{\rm frac}$ value, we compute the survival fraction both without and with considering relaxation to a Hernquist profile, and we choose the method that gives us the lesser value of survival fraction. Implementing this procedure in Python, we find that the algorithm selects the method without relaxation for low values of $E_{\rm frac}$. As $E_{\rm frac}$ is increased, the algorithm switches to the method with relaxation at some (unenforced) value of $E_{\rm frac}$.
	\begin{figure*}
		\includegraphics[width=\textwidth]{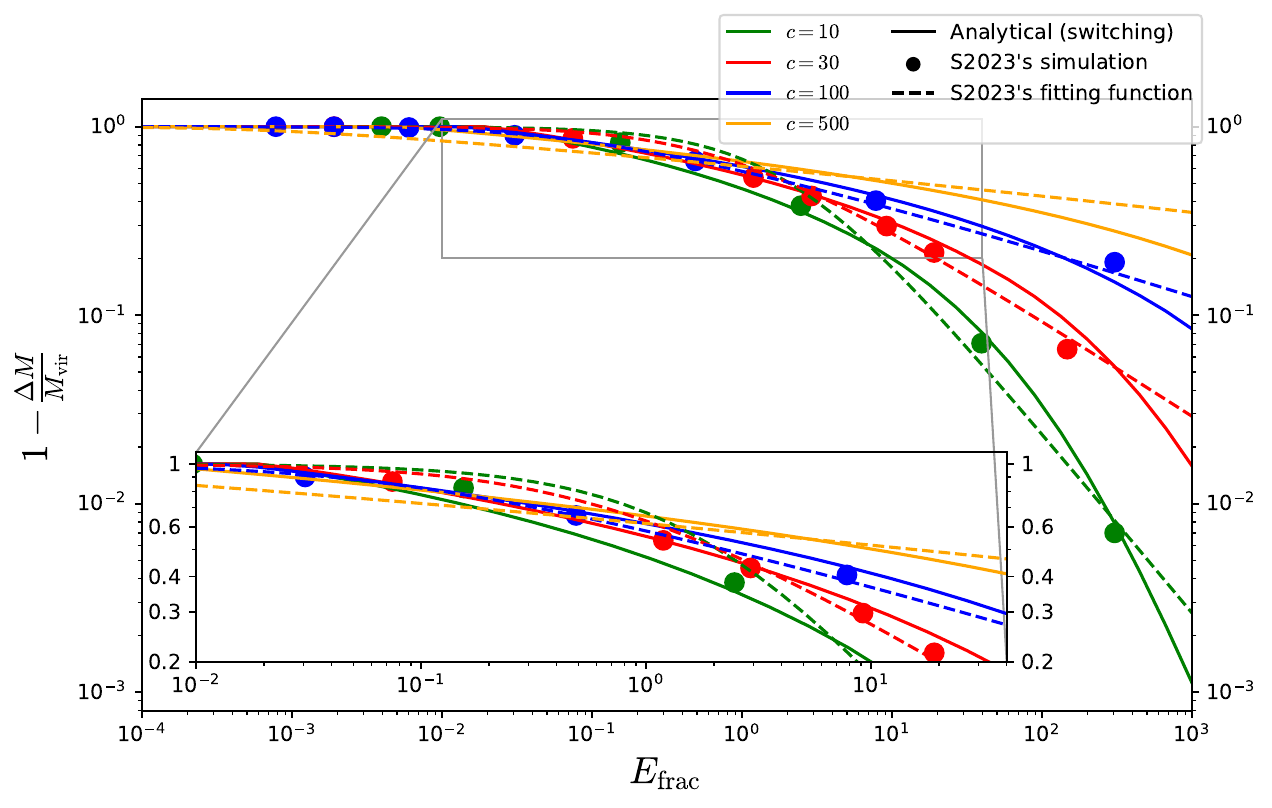}
		\caption{
			The same as Fig.~\ref{fig:SF vs Efrac for four values of concentration} except that the solid curves are the output of our analytical approach using the sequential stripping model of mass loss in the minihalo and incorporating the switching procedure.
		}
		\label{fig:SF vs Efrac - switching}
	\end{figure*}
	Fig.~\ref{fig:SF vs Efrac - switching} shows the survival fraction plotted against $E_{\rm frac}$ using the switching procedure.  The switching procedure provides a good match to the numerical data for all regimes of $E_{\rm frac}$ values considered.

	\begin{figure}
		\centering
		\includegraphics[width=\columnwidth]{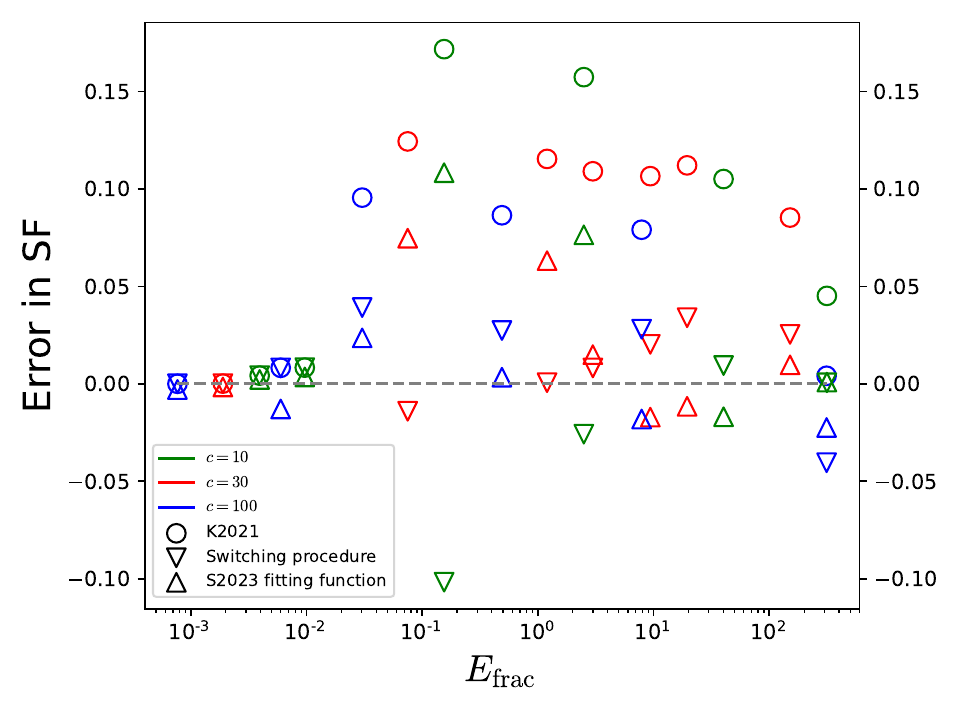}
		\caption{The error in a model is defined as (model \cg{$-$} data). Plotted are the errors in the output of Ref.~\cite{K2020}'s (K2021) analytical approach, our analytical
			switching procedure and S2023's empirical fitting functions. The numerical data is taken from S2023's N-body simulations. Errors are plotted for concentration parameters $c=10,30,100$.
		}
		\label{fig:error}
	\end{figure}
	\begin{figure}
		\centering
		\includegraphics[width=\columnwidth]{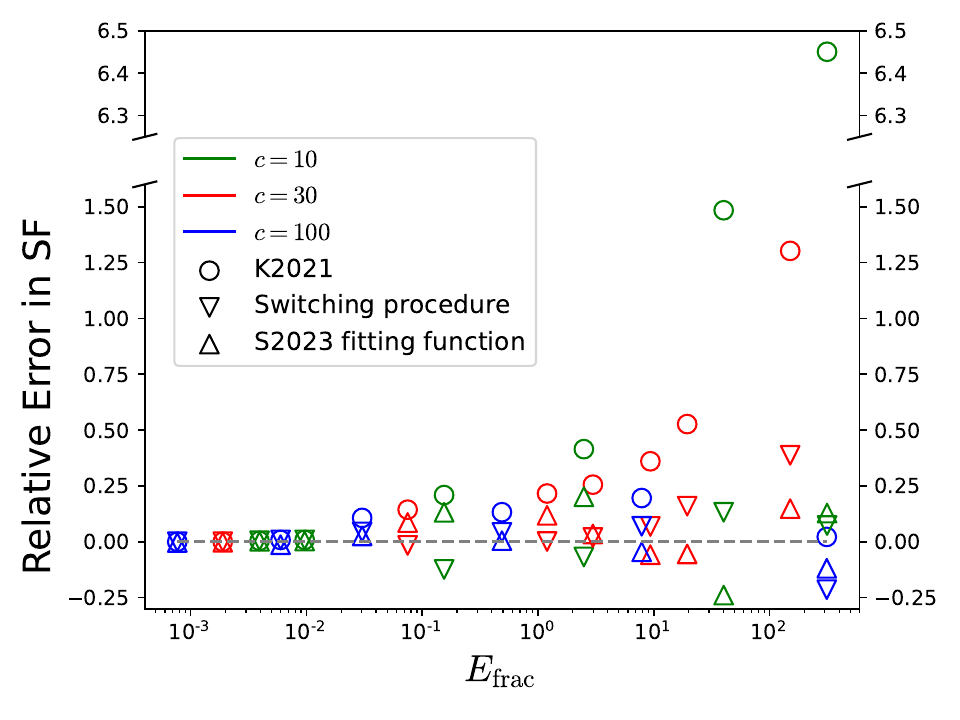}
		\caption{The same as Fig.~\ref{fig:error} except for the relative error, which is defined as (model \cg{$-$} data)/data, rather than the error.
		}
		\label{fig:relative error}
	\end{figure}
	Figs.~\ref{fig:error} and \ref{fig:relative error} show the error and relative error respectively in the survival fraction vs $E_{\rm frac}$ curves for K2021's approach, our analytical 
	switching procedure and S2023's fitting functions.

	As can be seen, the smallest errors are shown by the switching procedure and S2023's semi-analytic fitting functions. In some instances, the switching procedure outperforms S2023's fitting functions.

	\section{Multiple stellar encounters of an NFW minihalo}
	\label{sec:Multiple_stellar_encounters_of_an_NFW_minihalo}
	Here, we consider the scenario of multiple successive stellar interactions with an NFW minihalo. After each stellar interaction, the remnant minihalo is allowed to fully relax before the next stellar interaction is applied. S2023 states that after a stellar encounter, the NFW minihalo will relax to a broken power law profile with a $k$ dependent on the impact parameter of the stellar interaction. In general, $k \sim 3$. We will assume $k=3$ for our analytical calculations, i.e., a Hernquist profile. We also assume that when a Hernquist profile is perturbed, the remnant minihalo will relax to a Hernquist profile with a different concentration and overall mass.
	
	The NFW and Hernquist density profiles are two-parameter models. However, it turns out that we need only one piece of information to calculate the survival fraction (or mass loss fraction) due to a stellar encounter. Here, we use the concentration of the minihalo as that piece of information. As we will see later, the concentration of a two-parameter minihalo, like the NFW and Hernquist profiles, uniquely determines the scale density of that minihalo. So, we cannot compute the exact value of the scale radius as that would need a second piece of information. However, we can compute the ratio of the parent and child \cg{minihalos'} scale radii. This will allow us to calculate the scale density of the child minihalo. Knowing the scale density of the child minihalo gives us its concentration parameter. We then iterate this process to get the concentrations (and hence survival fractions) of successive generations of minihalos.
	
	We use the subscript $s$ to denote the unperturbed NFW minihalo. We use the numeral $n$ to denote the $n^{\text {th }}$ generation Hernquist minihalo, where the first-generation Hernquist minihalo is the child of the NFW minihalo and the $n^{\text {th }}$ generation Hernquist minihalo is the child of the $(n-1)^{\text {th }}$ generation Hernquist minihalo.
	
	To compare the survival fractions of successive generations of minihalos, we will define the survival fraction of any minihalo as
	
	\begin{equation}
		\text{SF}_n \equiv 1-\frac{\Delta M}{M_{\text {vir,s}}}    
	\end{equation}
	where $n = s, 1, 2, 3, \cdots$ and $\Delta M$ is the total mass lost within the virial radius of the unperturbed NFW minihalo. This means that our region of interest for the purposes of evaluating mass loss is always inside a 
	sphere of radius $r_{\rm vir,s}$.
	
	We start the procedure by specifying the concentration $c_{\rm s}$ of the NFW minihalo. To better understand how some of the equations in this section are derived, see Appendix~\ref{app:multiple stellar encounters}.
	
	\noindent\underline{Step I}: Compute the survival fraction of the NFW minihalo
	
	We use the switching procedure to find out the survival fraction of the NFW minihalo. We will restrict our incremental $E_{\rm frac}$ to be high enough such that the switching procedure forces the remnant minihalo to relax to a Hernquist profile. Knowing $c_{\rm s}$, we can calculate the scale density $\rho_{\rm s}$ of the NFW minihalo using Eq.~(\ref{eq:rho_s in terms of concentration for NFW profile}). Next we calculate $R_{\rm s}$ using Eq.~(\ref{eq:r_s definition}). We can then calculate the scale density $\rho_1$ of the first-generation Hernquist minihalo using Eq.~(\ref{eq:rho_1 final condition copy}). We can then calculate concentration $c_1$ of the first-generation Hernquist minihalo by solving for Eq.~(\ref{eq:relating concentration and scale radius for Hernquist profile}). The NFW minihalo's survival fraction is then given by Eq.~(\ref{eq:survival fraction of NFW minihalo incorporating relaxation}).

	\noindent\underline{Step II}: Compute the survival fraction of the $n^{\rm th}$ generation Hernquist minihalo ($n \geq 1$)
	
	In the $(n-1)^{\text{th}}$ step, we would have calculated the scale density $\rho_n$ and concentration $c_n$ of the $n^{\rm th}$ generation Hernquist minihalo. We next consider the conservation of mass condition for the transition from the $n^{\text{th}}$ to $(n+1)^{\text{th}}$ generation Hernquist minihalo.
	
	\begin{equation}\label{eq:mass condition for transition from n^th to (n+1)^th generation Hernquist minihalo}
		M_{\text {enc,}n}\left(x_n^*\right)-\Delta M_{x_n=0 \rightarrow x_n^*}=\lim _{x_{n+1} \rightarrow \infty} M_{\text {enc,}n+1}\left(x_{n+1}\right)
	\end{equation}
	Evaluating each of the terms in Eq.~(\ref{eq:mass condition for transition from n^th to (n+1)^th generation Hernquist minihalo}), it can be shown that the ratio of the scale radii of the $(n+1)^{\text{th}}$ to the $n^{\text{th}}$ generation Hernquist minihalos is (see Appendix~\ref{app:rn+1_by_rn and rho_n+1 derivation})
	
	\begin{equation}
		\frac{r_{n+1}}{r_n}=R_n
	\end{equation}
	where
	\begin{equation}\label{eq:R_n definition}
		R_n \equiv \sqrt{f_{\rm Hern}(c_n x_n^*)-8 \pi c_n^3 I_n}
	\end{equation}
	
	\begin{multline}
		I_n \equiv \int\limits_{x_n=0}^{x_n^*}\int\limits_{\epsilon=0}^{\vert\Delta\epsilon(x_n)\vert} \int\limits_{\psi_{\rm B}^\prime=0}^\epsilon \frac{1}{\sqrt{8}\pi^2} x_n^2 \sqrt{2(\psi_{\rm A}(x_n)-\epsilon)} \\
		\frac{1}{\sqrt{\epsilon - \psi_{\rm B}^\prime}} \frac{\mathrm{d}^2\varrho}{\mathrm{d}\psi_{\rm B}^{\prime^2}} \left(x_n^\prime(\psi_{\rm B}^\prime)\right) \mathrm{d}\mathrm{\psi_{\rm B}^\prime}\mathrm{d}\epsilon \mathrm{d}x_n
	\end{multline}
	We can now calculate the scale density $\rho_{n+1}$ of the $(n+1)^{\text{th}}$ generation Hernquist minihalo using the following relationship (see Appendix~\ref{app:rn+1_by_rn and rho_n+1 derivation}):
	
	\begin{equation}
		\rho_{n+1}=\frac{\rho_n}{R_n}
	\end{equation}
	Next, we compute the concentration $c_{n+1}$ of the $(n+1)^{\text{th}}$ generation Hernquist minihalo by adapting Eq.~(\ref{eq:relating concentration and scale radius for Hernquist profile}) and solving for $c_{n+1}$ in the expression below:
	
	\begin{equation}
		\frac{1}{2 c_{n+1}\left(1+c_{n+1}\right)^2}=\frac{200}{3} \frac{\rho_{\rm crit}}{\rho_{n+1}}
	\end{equation}
	The survival fraction of the $n^{\text{th}}$ generation Hernquist minihalo is given by (See Appendix~\ref{app:SF_n derivation})
	\begin{equation}
		\text{SF}_n=\frac{1}{2}\left(R_n R_{n-1} \ldots R_1 R_{\rm s}\right)^2 \frac{f_{\rm Hern}(c_{n+1} x_{n+1}^{r_{\text{vir,s}}})}{f_{\rm NFW}(c_{\rm s})}
	\end{equation}
	where
	\begin{equation}
		x_{n+1}^{r_{\text{vir,s}}}=\frac{c_{\rm s}}{c_{n+1}} \frac{1}{R_n R_{n-1} \cdots R_1 R_{\rm s}}
	\end{equation}
	Step II is applied for $n=1$, then $n=2$, and so on. This completes the theoretical procedure for computing survival fractions for multiple stellar encounters of an NFW minihalo.
	\begin{figure}
		\includegraphics[width=\columnwidth]{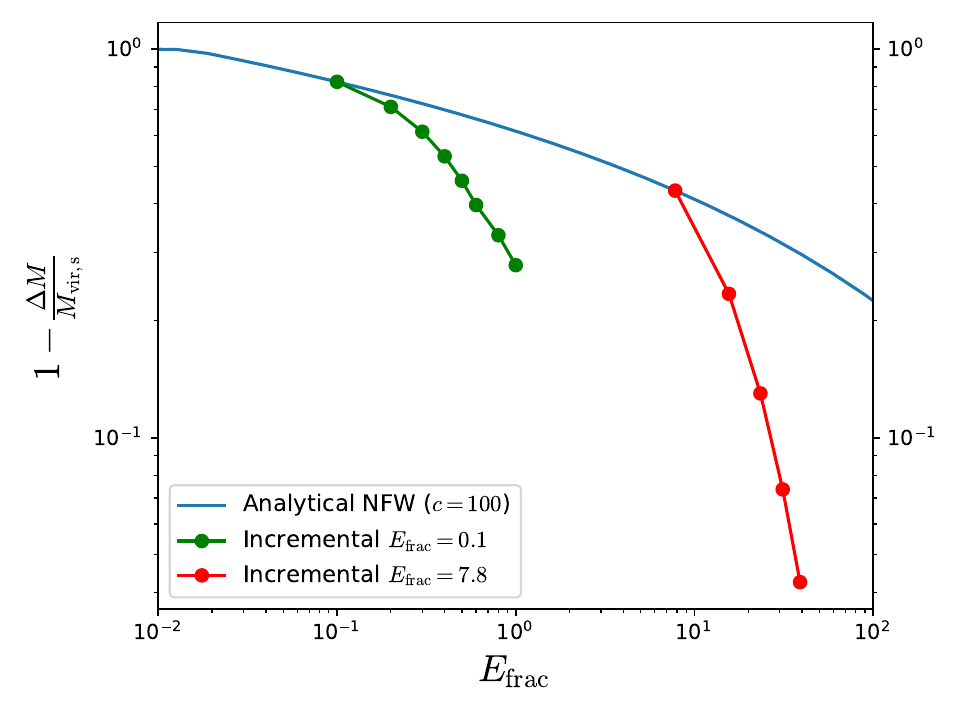}
		\caption{The survival fraction of an NFW minihalo is plotted against the total normalized injected energy $E_{\rm frac}$. The blue curve shows the case of a single stellar encounter. The green (red) curves represent the multiple stellar encounter scenario where an incremental $E_{\rm frac} = 0.1 (7.8)$ is used repeatedly. When the incremental $E_{\rm frac}$ is higher (red curve), the deviation from the single energy injection case (blue curve) is higher as well.}
		\label{fig:survival fractions for multiple encounters - fixed Efrac}
	\end{figure}
	We now set $c_{\rm s} = 100$ and evaluate the survival fractions of the NFW minihalo and successive generations of Hernquist minihalos using the above mentioned procedure. Fig.~\ref{fig:survival fractions for multiple encounters - fixed Efrac} shows the survival fractions resulting from multiple energy injections into the minihalo due to successive stellar encounters. The blue curve represents the survival fraction vs $E_{\rm frac}$ due to single stellar encounter scenarios. The green/red curves represent multiple encounters, each characterized by an incremental $E_{\rm frac} = 0.1 / 7.8$, respectively. However, the green curve's last two incremental energy injections are characterized by an $E_{\rm frac} = 0.2$.
	
	It can be seen that multiple encounters generate more mass loss than a single-shot encounter case of the same cumulative energy injection $E_{\rm frac}$ \cg{because of the relaxation occurring in between stellar encounters}.
	
	In Fig.~\ref{fig:survival fractions for multiple encounters - fixed Efrac}, we have fixed the incremental $E_{\rm frac}$ between successive encounters. According to S2023, for a large enough impact parameter,
	
	\begin{equation}\label{eq:Efrac expression in terms of encounter parameters}
		E_{\rm frac} = \frac{\alpha^2}{\pi \gamma} \frac{G M_\star^2}{V^2 b^4} \frac{1}{\bar{\rho}_{\rm vir}}
	\end{equation}
	where $M_\star$ is the mass of the perturbing star, $V$ is the relative velocity of the star, and $\bar{\rho}_{\rm vir}$ is the average density of the minihalo within its virial radius. $\alpha^2$ and $\gamma$ are functions of $c$, which depends on the density profile of the minihalo - specifically, it depends on the scale density of the minihalo. Between successive encounters of the minihalo, its density profile changes. Thus, the ratio $\frac{\alpha^2}{\gamma}$ changes. Assuming we fix $M_\star$, $V$ and $\bar{\rho}_{\rm vir}$ between encounters (recall $\bar{\rho}_{\rm vir}=200 \rho_{\text{crit}}$, so it is fixed), fixing $E_{\rm frac}$ necessarily means that the impact parameter $b$ changes between successive encounters.
	\begin{figure}
		\includegraphics[width=\columnwidth]{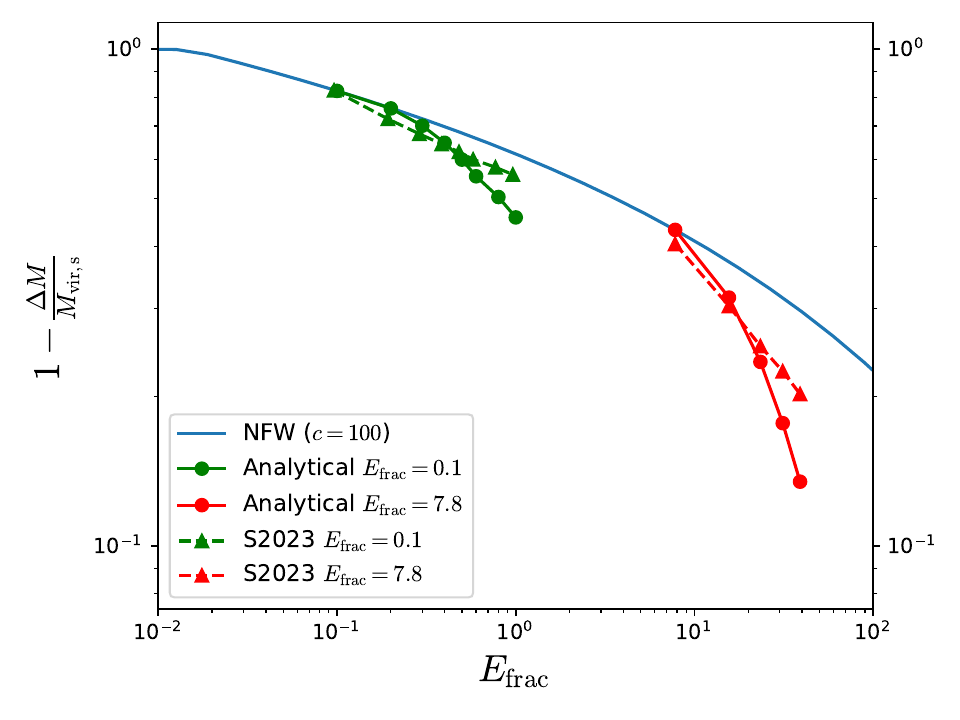}
		\caption{The survival fraction of an NFW minihalo is plotted against the total normalized injected energy $E_{\rm frac}$. The blue curve shows the case of a single stellar encounter. The green (red) curves represent the multiple stellar encounter scenario where an incremental $E_{\rm frac} = 0.1 (7.8)$ is used repeatedly. The solid curves with circular markers represent the output of our analytical method, while the dashed curves with triangular markers represent the output of S2023's N-body simulations.}
		\label{fig:survival fractions for multiple encounters - fixed Efrac and fixed b}
	\end{figure}
	In Fig.~9 of S2023, they have a graph where they use their N-body results to evaluate the survival fractions of an NFW minihalo for a multiple stellar encounter scenario. We would like to compare our results with that of S2023. They use incremental $E_{\rm frac} = 0.1, 0.1, 0.1, 0.1, 0.1, 0.1, 0.2, 0.2$ in one case and $E_{\rm frac} = 7.8, 7.8, 7.8, 7.8, 7.8$ in a second case. We will also use these values of incremental $E_{\rm frac}$ in our analytical calculations. S2023 evaluates $b$ from Eq.~(\ref{eq:Efrac expression in terms of encounter parameters}) for an NFW profile, given knowledge of the concentration of the NFW minihalo and the $E_{\rm frac}$ value. S2023 then fixes $b$ between successive encounters while simultaneously assuming that $E_{\rm frac}$ is held constant $(0.1 / 0.2 / 7.8)$~\cite{Shen_personal_communication_2023}.
	This approach is inaccurate since the density profile of the minihalo changes between encounters, thus changing the ratio $\frac{\alpha^2}{\gamma}$. Thus, both $E_{\rm frac}$ and $b$ can't be fixed between encounters. As we mentioned, S2023 fixes $b$ between successive encounters. To compare the output of our analytical procedure to S2023's results, we emulate this by first evaluating $b$ from Eq.~(\ref{eq:Efrac expression in terms of encounter parameters}) for the NFW profile, with a known $E_{\rm frac}$ $(0.1 / 0.2 / 7.8)$ and then fix that value of $b$ and evaluate the actual $E_{\rm frac}$ using Eq.~(\ref{eq:Efrac expression in terms of encounter parameters}) for the remaining encounters. Using the actual value of $E_{\rm frac}$ for each encounter, we analytically compute the survival fraction for that encounter. However, when plotting the results in a graph, we use $E_{\rm frac} = 0.1/0.2/7.8$ for each encounter instead of the actual value of $E_{\rm frac}$. This allows us to compare the output of our analytical procedure to the numerical simulations of S2023. Fig.~\ref{fig:survival fractions for multiple encounters - fixed Efrac and fixed b} shows the corresponding results. The blue curve represents the single encounter case. The green (red) curves represent the multiple encounter scenario with incremental $E_{\rm frac}=0.1 (7.8)$. The solid curves with circular markers represent our analytical method, while the dashed curves with triangular markers represent S2023's numerical simulations. There is a fair amount of agreement between our results and S2023's. However, the slight deviation in the two results is likely because, as is apparent from Fig.~\ref{fig:SF vs Efrac - switching}, our analytical method of computing survival fractions gives a slightly different answer compared to S2023's numerical simulations. Moreover, we assume that the successive generations of minihalos have Hernquist density profiles. However, in reality, those minihalos will have a broken power law profile with the $k$ parameter close to, but not exactly, 3 (which would be the Hernquist profile).
	
	\section{Empirical method of accounting for multiple encounters}
	Inspired by ref.~\cite{Stucker2023}, we propose the following empirical method 
	for evaluating the effective single energy injection from multiple stellar encounters:
	\begin{equation}
		\label{eq:multiple encounters}
		E_{\rm frac,eff} = \left(  \sum_i E_{{\rm frac},i}^{p/2} \right)^{2/p}
	\end{equation}
	where $p$ is a parameter to be determined. 
	For $p=2$, the value of $E_{\rm frac,eff} $ would correspond to the sum of all individual $E_{{\rm frac},i}$. For $p<2$, multiple energy injections would have an enhanced effect. Only the strongest energy injection would matter for $p \rightarrow$ $\infty$. 
	
	Ref.~\cite{Stucker2023} finds that $p=1.2$ for a prompt cusp which has a density that differs from the NFW form, following a steep $r^{-1.5}$ density profile between outer boundary set by the curvature of the initial density peak and inner core determined by the physical nature of the dark matter. 
	S2023 effectively assumed $p=2$. As our results in the previous section indicate, $p<2$ would provide a better fit.   
	A similar finding was obtained by ref.~\cite{Delos:2019tsl}. However,
	due to the different methods and assumptions used there, it is difficult to make a precise comparison between our results.
	
\begin{figure}
		\includegraphics[width=\columnwidth]{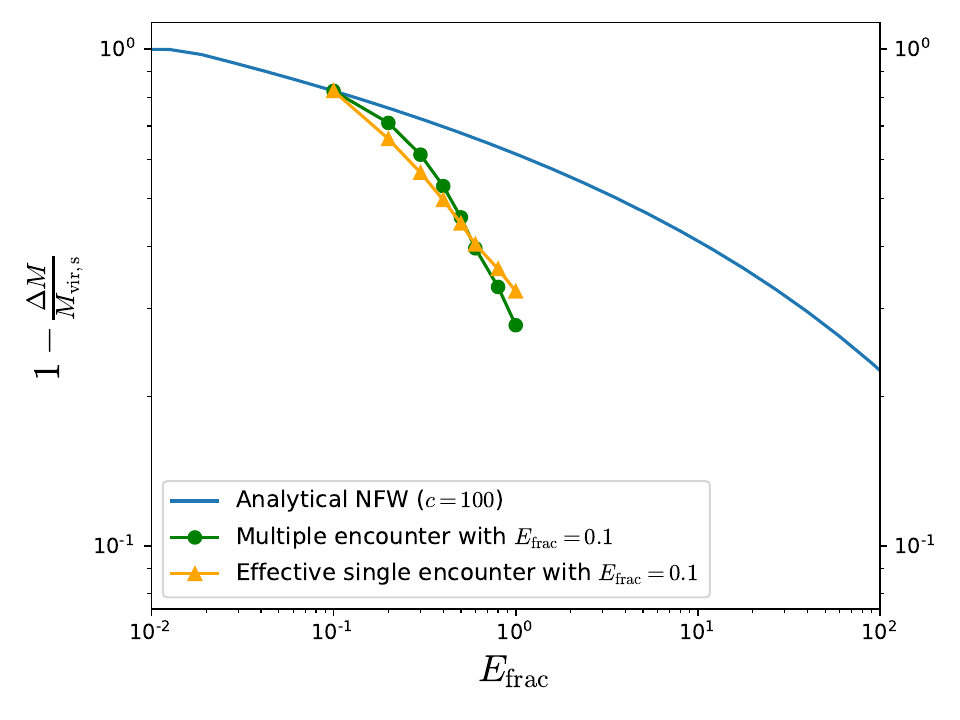}	\includegraphics[width=\columnwidth]{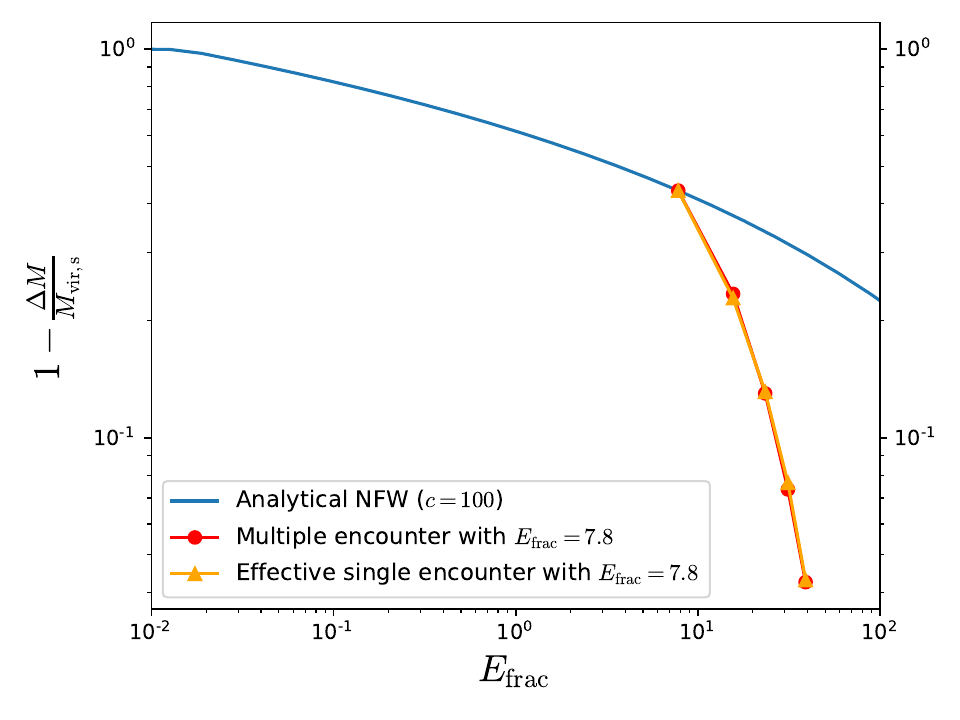}
		\caption{The survival fraction (SF) of an NFW minihalo is plotted against the total normalized energy energy injection $E_{\rm frac}$. The blue curve shows the case of a single stellar encounter. The green (red) curves represent the multiple stellar encounter scenario where an incremental $E_{\rm frac} = 0.1 (7.8)$ is used repeatedly. In the top and bottom panels, the orange curve is the best fit $p$ using Eq. \ref{eq:multiple encounters} for individual energy injections of $E_{\rm frac}=0.1$  and 7.8, respectively. The respective best-fit values of $p$ were 0.8 and 0.6. 
			\label{fig:survival fractions for multiple encounters - fixed Efrac - best fit p}}
	\end{figure}

	We performed a least squares fit to find the optimal value of $p$ for the cases we investigated in Figs.~\ref{fig:survival fractions for multiple encounters - fixed Efrac}. The results are shown in Fig.~\ref{fig:survival fractions for multiple encounters - fixed Efrac - best fit p}. Our best-fit values of $p=0.8$ and $p=0.6$ 
	are consistent with our earlier findings that successive encounters are more destructive than a single effective encounter, with the effective \cg{fractional} energy equal to the sum of the effective \cg{fractional} energies of each actual encounter.

	\section{Discussion and Conclusions}
	\label{sec:Discussion_and_Conclusions}

	In extending the K2021 method for evaluating minihalo mass loss due to stellar encounters, we have introduced a sequential stripping model. This model conceptually divides the minihalo into infinitesimally thin shells, sequentially focusing from the outermost to the innermost. However, this sequential approach does not imply a temporal sequence in the stripping process. Instead, it is a methodological tool for analysis. In reality, during a stellar encounter, the stripping of shells occurs simultaneously, regardless of their distance from the minihalo center.

	Building on this, it is crucial to note that our model also accounts for changes in the minihalo's profile after a stellar encounter. This aspect becomes significant if the energy injection during the encounter is sufficiently large, altering the minihalo's structure and dynamics.
	
	As shown in  Figs.~\ref{fig:error} and \ref{fig:relative error}, our new method provides a significantly better fit to S2023's N-body simulation results
	compared to the K2021 method. This is particularly noticeable for $E_{\rm frac}\gtrsim 10^{-2}$, which is relevant for modeling the mass loss incurred by multiple passes through the Milky Way disk (S2023).

	A significant finding of our research centers on the treatment of minihalos undergoing multiple stellar encounters. 
	Note that this is for the case where the minihalo has had time to stabilize after each encounter.  As discussed in S2023, for example, this scenario \cg{may be} appropriate for successive cases of the minihalo passing through the Galactic disk. Although many encounters may occur during a single passing through of the disk, these will be in such rapid succession that they can be considered one encounter where the \cg{fractional} energies have been summed up to give the effective \cg{fractional} energy of one encounter.
	
	Contrary to the results presented by S2023, our analysis suggests that sequential stellar interactions lead to a more pronounced mass loss in minihalos when they are allowed to fully relax between encounters. This finding will have implications for our understanding of minihalo survival and evolution in dense stellar environments. S2023 found that by $z=0$ at the Solar System location, around 60\% of the mass 
	in minihalos has survived stellar disruption from the Milky Way disk. However, as we have shown, they assume the fractional energy injections for each passage through the disk can be added to make a single effective fractional energy injection. Our results indicate that this underestimates the effects of sequential energy injections. In future work, we would like to 
	use our new method of evaluating the impact of sequential stellar encounters to estimate the mass lost
	by minihalos in the Milky Way. 

 All code used in this article can be found at the following online repository: \url{https://github.com/ian-dsouza/axion-minihalos}.
	
	\begin{acknowledgements} \cg{Ian DSouza is funded by the University of Canterbury Doctoral Scholarship.}
		We thank Ciaran O'Hare for helpful discussions and Xuejian (Jacob) Shen for supplying some of the data points from S2023's N-body simulations.
	\end{acknowledgements}

	\bibliography{axion_mini_cluster_disruption.bib}
	
	\appendix
	
	\section{Computing the expression for the normalized relative potential of an untruncated NFW minihalo}\label{app:psi for untruncated NFW}
	The Newtonian gravitational potential for an untruncated NFW density profile is Ref.~\cite[equation 2.67]{binneyTremaine}
	\begin{equation}\label{eq:Phi for untruncated NFW}
		\Phi(r) = -4\pi G\rho_{\rm s}r_{\rm s}^3 \frac{\ln\left(1+\frac{r}{r_{\rm s}}\right)}{r}
	\end{equation}
	Making the substitutions $\frac{r}{r_{\rm s}} = \frac{xr_{\rm vir}}{r_{\rm s}} = cx$ and $r_{\rm s} = \frac{r_{vir}}{c}$ in Eq.~(\ref{eq:Phi for untruncated NFW})
	\begin{align}\label{eq:Phi(x) intermediate}
		\Phi(x) &= -4\pi G\rho_{\rm s}\frac{r_{\rm vir}^3}{c^3} \frac{\ln\left(1+cx\right)}{xr_{\rm vir}} \nonumber\\
		&= -\Psi_0 \times \frac{4\pi \rho_{\rm s}r_{\rm vir}^3}{M_{\rm vir}c^3}\frac{\ln\left(1+cx\right)}{x} 
	\end{align}
	where $\Psi_0 \equiv \frac{GM_{\rm vir}}{r_{\rm vir}}$
	
	Substituting for $M_{\rm vir}$ from Eq.~(\ref{eq:M_vir}) in Eq.~(\ref{eq:Phi(x) intermediate})
	\begin{equation}\label{eq:Phi intermediate 2}
		\Phi(x) = -\Psi_0  \frac{1}{f_{\rm NFW}(c)}\frac{\ln\left(1+cx\right)}{x} 
	\end{equation}
	Let the relative potential be $\Psi \equiv -\Phi$ and the normalized relative potential be $\psi \equiv \frac{\Psi}{\Psi_0}$. $\implies \psi = -\frac{\Phi}{\Psi_0}$. From Eq.~(\ref{eq:Phi intermediate 2}), this implies that the normalized relative potential is
	\begin{equation}
		\psi(x) =   \frac{1}{f_{\rm NFW}(c)}\frac{\ln\left(1+cx\right)}{x} 
	\end{equation}
	\section{Computing the dimensionless expression for the survival fraction of an NFW minihalo}\label{app:survival fraction expression for NFW minihalo}
	We start with Eq.~(\ref{eq:mass loss NFW}). Substituting with $r = xr_{\rm vir}$, $\Psi = \psi\Psi_0$, $\varepsilon = \epsilon\Psi_0$, $\vert\Delta\varepsilon\vert = \vert\Delta\epsilon\vert \Psi_0$ and $f = \frac{\rho_{\rm s}}{\Psi_0^{3/2}} \hat{f}$, we get
	\begin{multline}\label{eq:Delta M dimensionless}
		\Delta M = 16\pi^2\rho_{\rm s}r_{\rm vir}^3 \int\limits_{x=0}^1 \mathrm{d}x\ x^2 \int\limits_{\epsilon=0}^{\min\left[\vert\Delta\epsilon(x)\vert, \psi(x)\right]} \mathrm{d}\epsilon\\
		\sqrt{2(\psi(x)-\epsilon)} \hat{f}(\epsilon)
	\end{multline}
	From Eq.~(\ref{eq:M_vir}),
	\begin{equation}\label{eq:M_vir repeated}
		M_{\rm vir} = 4\pi\rho_{\rm s}r_{\rm vir}^3 \frac{f_{\rm NFW}(c)}{c^3}
	\end{equation}
	From Eqs.~(\ref{eq:Delta M dimensionless}) and (\ref{eq:M_vir repeated}), it follows that the survival fraction is
	\begin{align}\label{eq:surival fraction expression dimensionless double integral REPEATED}
		\text{SF} &\equiv 1 - \frac{\Delta M}{M_{\rm vir}} \nonumber\\
		&= 1 - \frac{4\pi c^3}{f_{\rm NFW}(c)}\int\limits_{x=0}^1\mathrm{d}x\ x^2\int\limits_{\epsilon=0}^{\min[\vert\Delta\epsilon(x)\vert, \psi(x)]} \mathrm{d}\epsilon\ \hat{f}(\epsilon)\nonumber\\
		&\ \ \ \ \ \ \ \ \ \ \ \  \ \ \ \ \ \ \ \ \ \ \ \ \ \ \ \ \ \ \ \ \ \ \ \sqrt{2(\psi(x)-\epsilon)}
	\end{align}
	\section{Computing the mass of the minihalo in a shell of finite thickness}\label{app:Computing mass of minihalo}
	K2021 gives the phase space distribution of dark matter particles in a minihalo as:
	\begin{equation}\label{eq:phase space distribution definition}
		f(\varepsilon) \equiv \frac{\mathrm{d}M}{\mathrm{d}^3\vec{r}\mathrm{d}^3\vec{v}}
	\end{equation}
	\begin{equation}\label{eq:varepsilon definition}
		\varepsilon \equiv \Psi(r) - \frac{v^2}{2}
	\end{equation}
	where $\varepsilon$ is called the specific relative (total) energy, $\vec{r}$ is the radius vector associated with a dark matter particle, $\vec{v}$ is the corresponding velocity vector, $\mathrm{d}M$ is the mass present in the phase space volume of $\mathrm{d}^3\vec{r}\mathrm{d}^3\vec{v}$. Assuming spherical symmetry in physical space and velocity space,
	\begin{equation}\label{eq:dM}
		\mathrm{d}M = 16\pi^2 f(\varepsilon) r^2v^2\mathrm{d}v\mathrm{d}r
	\end{equation}
	To find the mass of the minihalo between $r=r_1$ and $r=r_2$, we integrate Eq.~(\ref{eq:dM})
	\begin{equation}\label{eq:DeltaM equals integral of dM}
		\cg{\delta} M = 16\pi^2 \int\limits_{r=r_1}^{r_2} \int\limits_{v=0}^{v_{\rm max}(r)} f(\varepsilon) r^2v^2\mathrm{d}v\mathrm{d}r
	\end{equation}
	While performing the $v$-integral, $r$ remains fixed. Thus, differentiating Eq.~(\ref{eq:varepsilon definition})
	\begin{equation}\label{eq:dVarepsilon}
		\mathrm{d}\varepsilon = -v\mathrm{d}v
	\end{equation}
	Moreover, using Eq.~(\ref{eq:varepsilon definition}), solving for $v$
	\begin{equation}\label{eq:v solved}
		v = \sqrt{2(\Psi(r) - \varepsilon)}
	\end{equation}
	Also, we note that when $v = v_{\rm max}(r)$, $\varepsilon=0$ from Eq.~(\ref{eq:varepsilon definition}) since $\varepsilon$ cannot be negative for a dark matter particle bound to the minihalo.
	
	Using Eqs.~(\ref{eq:dVarepsilon}) and (\ref{eq:v solved}), Eq.~(\ref{eq:DeltaM equals integral of dM}) becomes
	\begin{equation}\label{eq:Delta M final equation}
		\cg{\delta M} = 16\pi^2 \int\limits_{r=r_1}^{r_2} \int\limits_{\varepsilon=0}^{\Psi(r)} f(\varepsilon) r^2\sqrt{2(\Psi(r) - \varepsilon)}\mathrm{d}\varepsilon\mathrm{d}r
	\end{equation}
	Thus, when the upper limit of the $\varepsilon$-integral is $\Psi(r)$, Eq.~(\ref{eq:Delta M final equation}) gives the mass of the minihalo between the radii $r=r_1$ and $r=r_2$.
	
	\section{Evaluating the survival fraction using the sequential stripping model}
	\label{app:Evaluating the survival fraction using the sequential stripping model}
	Eq.~(\ref{eq:cross over radius condition}) can be used to split the triple integral in Eq.~(\ref{eq:surival fraction expression dimensionless triple integral}) into two triple integrals as follows:
	\begin{multline}\label{eq:surival fraction expression dimensionless with crossover radius}
		\text{SF} = 1 - 
		\frac{4\pi c^3}{f_{\rm NFW}(c)}\biggr[ 
		\int\limits_{x=0}^{\min[x^*, 1]} \int\limits_{\epsilon=0}^{\vert\Delta\epsilon(x)\vert} \  \int\limits_{\psi^\prime=0}^\epsilon \\
		x^2\frac{1}{\sqrt{8}\pi^2}\ \frac{1}{\sqrt{\epsilon - \psi^\prime}}
		\frac{\mathrm{d}^2\varrho}{\mathrm{d}\psi^{\prime^2}}  \sqrt{2(\psi(x)-\epsilon)} \ \mathrm{d}\psi^\prime\ \mathrm{d}\epsilon\ \mathrm{d}x\\
		+    \int\limits_{x=\min[x^*, 1]}^1 \int\limits_{\epsilon=0}^{\psi(x)} \  \int\limits_{\psi^\prime=0}^\epsilon \\
		x^2\frac{1}{\sqrt{8}\pi^2}
		\frac{1}{\sqrt{\epsilon - \psi^\prime}} \frac{\mathrm{d}^2\varrho}{\mathrm{d}\psi^{\prime^2}} \sqrt{2(\psi(x)-\epsilon)} \ \mathrm{d}\psi^\prime\ \mathrm{d}\epsilon\ \mathrm{d}x\biggr]
	\end{multline}
	Since we are only interested in the mass loss within the virial radius, a $\min[x^*, 1]$ term is introduced in the limits of the $x$-integral in Eq.~(\ref{eq:surival fraction expression dimensionless with crossover radius}) to account for the case when $x^* > 1$, where $x=1$ represents the virial radius.

	Let's rewrite Eq.~(\ref{eq:surival fraction expression dimensionless with crossover radius}) as follows:
	\begin{equation}\label{eq:survival fraction expression decomposed}
		\text{SF} = 1 - \text{prefactor} \times [I_{\rm A} + I_{\rm B}]
	\end{equation}
	where
	\begin{equation}\label{eq:prefactor}
		\text{prefactor} \equiv \frac{4\pi c^3}{f_{\rm NFW}(c)}
	\end{equation}
	\begin{multline}\label{eq:I1}
		I_{\rm A} \equiv \int\limits_{x=0}^{\min[x^*, 1]} \int\limits_{\epsilon=0}^{\vert\Delta\epsilon(x)\vert} \  \int\limits_{\psi^\prime=0}^\epsilon x^2\frac{1}{\sqrt{8}\pi^2}\ \frac{1}{\sqrt{\epsilon - \psi^\prime}} \frac{\mathrm{d}^2\varrho}{\mathrm{d}\psi^{\prime^2}}\\ \sqrt{2(\psi(x)-\epsilon)} \ \mathrm{d}\psi^\prime\ \mathrm{d}\epsilon\ \mathrm{d}x\
	\end{multline}
	\begin{multline}\label{eq:I2}
		I_{\rm B} \equiv \int\limits_{x=\min[x^*, 1]}^1 \int\limits_{\epsilon=0}^{\psi(x)} \  \int\limits_{\psi^\prime=0}^\epsilon x^2\frac{1}{\sqrt{8}\pi^2}\ \frac{1}{\sqrt{\epsilon - \psi^\prime}} \frac{\mathrm{d}^2\varrho}{\mathrm{d}\psi^{\prime^2}} \\ \sqrt{2(\psi(x)-\epsilon)} \ \mathrm{d}\psi^\prime\ \mathrm{d}\epsilon\ \mathrm{d}x
	\end{multline}
	For illustration purposes, let's assume that $x^*<1$. Then, according to Eq.~(\ref{eq:I2}), $I_{\rm B}$ is non-zero. Here, ``prefactor $\times\ I_{\rm B}$" represents mass loss in the region from $x=x^*$ to $x=1$. It is important to note that when the upper limit of the $\epsilon$-integral is $\psi(x)$, then the triple integral calculates the mass of the minihalo between the lower and upper limits of the $x$-integral (see Appendix \ref{app:Computing mass of minihalo}). This implies a total mass loss in the region $x\in\left[x^*,1\right]$. To be exact, ``prefactor $\times\ I_{\rm B}$" is the mass loss fraction (relative to the virial mass) in the region $x\in\left[x^*,1\right]$. Then, ``1 - prefactor $\times\ I_{\rm B}$" represents the mass fraction of the region $x\in[0, x^*]$, since all mass in the region $x\in[x^*, 1]$ is lost due to stellar interaction. Thus,
	\begin{equation}\label{eq:mass fraction x < x^*}
		1 - \text{prefactor} \times I_{\rm B} = \text{mass fraction}_{x < \min[x^*,1]}
	\end{equation}
	Thus, Eq.~(\ref{eq:survival fraction expression decomposed}) can be written as:
	\begin{equation}\label{eq:SF expression in terms of mass fraction x < x^* and I_A}
		\text{SF} = \text{mass fraction}_{x < \min[x^*,1]} - \text{prefactor}\times I_{\rm A}
	\end{equation}
	According to Fig.~\ref{fig:psi crossing over deltaEpsilon}, for $x<x^*$, $\vert\Delta\epsilon(x)\vert < \psi(x)$. Since the upper limit of the $\epsilon$-integral in Eq.~(\ref{eq:I1}) is $\vert\Delta\epsilon(x)\vert$, ``prefactor $\times\ I_{\rm A}$" represents a partial mass loss occurring in the region $x\in[0, x^*]$.
	
	We now have to mathematically evaluate the two terms of Eq.~(\ref{eq:SF expression in terms of mass fraction x < x^* and I_A}). Since ``mass fraction$_{x < \min[x^*,1]}$" represents the mass of the minihalo in the region $x\in\left[0, \min[x^*,1]\right]$, we can calculate it by simply integrating the NFW density profile from $x=0$ to $x=\min[x^*,1]$, and dividing the result by the virial mass which is a known expression.
	\begin{multline}\label{eq:expression for mass fraction x < x^*, in terms of r}
		\text{Mass fraction}_{x < \min[x^*,1]} = \frac{1}{M_{\rm vir}} \int\limits_{r=0}^{\min[r^*, r_{\rm vir}]} \frac{\rho_{\rm s}}{\frac{r}{r_{\rm s}}\left(1 + \frac{r}{r_{\rm s}}\right)^2}\\
		\times 4\pi r^2\mathrm{d}r
	\end{multline}
	where $r^*$ is the crossover radius defined as $r^* \equiv x^*r_{\rm vir}$.
	
	It can be shown that Eq.~(\ref{eq:expression for mass fraction x < x^*, in terms of r}) reduces to (see Appendix \ref{app:Computing mass fraction below crossover radius}):
	\begin{equation}\label{eq:expression for mass fraction x < x^*, in terms of x}
		\text{Mass fraction}_{x < \min[x^*,1]} = \frac{c^2}{f_{\rm NFW}(c)} \int\limits_{x=0}^{\min[x^*, 1]} \frac{x}{\left(1 + cx\right)^2} \mathrm{d}x
	\end{equation}
	
	For $x>x^*$, the resulting relative potential $\psi_{\rm B}(x)$ is given by (see Appendix (\ref{apppsiforxgreaterthanxstarsequentialstrippingmodel})),
	\begin{align}\label{eq:psi2(x)}
		\psi_{\rm B}(x) = \frac{1}{f_{\rm NFW}(c)} \left[ \frac{\ln(1+cx)}{x} - \frac{c}{1+cx} \right] && ,x>x^*
	\end{align}
	To compute the expression for the relative potential in the region $x<x^*$, while taking the dark matter particle from $x$ to infinity, we assume that all matter is intact from normalized radius $x$ to $x^*$, and all matter is already stripped off in the region $x>x^*$. The resulting relative potential $\psi_{\rm A}(x)$ is given by (see Appendix (\ref{app:psi(x) for x < x^* - sequential stripping model})),
	\begin{align}\label{eq:psi1(x)}
		\psi_{\rm A}(x) = \frac{1}{f_{\rm NFW}(c)} \left[ \frac{\ln(1+cx)}{x} - \frac{c}{1+cx^*} \right] && ,x<x^*
	\end{align}
	It is important to note that $\psi(x)$ is continuous at $x=x^*$ and hence $x^*$ can be easily evaluated using Eq.~(\ref{eq:cross over radius condition}).
	
	To evaluate the mass loss in the region $x<x^*$, we have to evaluate ``prefactor $\times\ I_{\rm A}$'' in Eq.~(\ref{eq:SF expression in terms of mass fraction x < x^* and I_A}). In the expression for $I_{\rm A}$, as given by Eq.~(\ref{eq:I1}), the $\sqrt{2(\psi(x)-\epsilon)}$ term becomes $\sqrt{2(\psi_{\rm A}(x)-\epsilon)}$ because the $x$-integral in Eq.~(\ref{eq:I1}) ranges from $x=0$ to $x=x^*$. Thus, $\psi_{\rm A}(x)$ from Eq.~(\ref{eq:psi1(x)}) applies here.
	
	In Eq.~(\ref{eq:I1}), there is also a term $\frac{\mathrm{d}^2\varrho}{\mathrm{d}\psi^{\prime^2}}$. Here , we have to ascertain if it is $\psi_{\rm A}(x)$ or $\psi_{\rm B}(x)$ that we differentiate $\varrho$ by. To find out, note that from the $\psi^\prime$-integral in Eq.~(\ref{eq:I1}), we have
	\begin{equation}\label{eq:psi_prime condition}
		0 \leq \psi^\prime \leq \epsilon
	\end{equation}
	From the $\epsilon$-integral in Eq.~(\ref{eq:I1}), we have
	\begin{equation}\label{eq:epsilon condition}
		\epsilon \leq \vert\Delta\epsilon(x)\vert\bigg\vert_{x\leq x^*}
	\end{equation}
	From Eqs.~(\ref{eq:normalized energy injected per unit mass (part 2) as a function of x}) and (\ref{eq:cross over radius condition}) we see that 
	\begin{equation}\label{eq:Delat_epsilon(x) condition}
		\vert\Delta\epsilon(x)\vert\bigg\vert_{x\leq x^*} \leq \vert\Delta\epsilon(x^*)\vert = \psi(x^*)
	\end{equation}
	This can also be seen in Fig.~\ref{fig:psi crossing over deltaEpsilon} for the $c=10$ case.
	
	From Eqs.~(\ref{eq:psi_prime condition}), (\ref{eq:epsilon condition}) and (\ref{eq:Delat_epsilon(x) condition}), it follows that
	\begin{equation}\label{eq:psi_prime final condition}
		\psi^\prime \leq \psi(x^*)
	\end{equation}
	Fig.~\ref{fig:psi crossing over deltaEpsilon} then says that $\psi^\prime$ is in the region $x\geq x^*$. Thus, we must use $\psi_{\rm B}^\prime(x^\prime)$ to differentiate $\varrho$ in $\frac{\mathrm{d}^2\varrho}{\mathrm{d}\psi^{\prime^2}}$. Thus, $I_{\rm A}$ (from Eq.~(\ref{eq:I1})) can be rewritten as
	\begin{multline}\label{eq:I1 updated to sequential stripping model}
		I_{\rm A} \equiv \int\limits_{x=0}^{\min[x^*, 1]} \int\limits_{\epsilon=0}^{\vert\Delta\epsilon(x)\vert} \  \int\limits_{\psi_{\rm B}^\prime=0}^\epsilon x^2\frac{1}{\sqrt{8}\pi^2}\ \frac{1}{\sqrt{\epsilon - \psi_{\rm B}^\prime}} \frac{\mathrm{d}^2\varrho}{\mathrm{d}\psi_{\rm B}^{\prime^2}}\left(x^\prime\left(\psi_{\rm B}^\prime\right)\right)\\ \sqrt{2\left(\psi_{\rm A}(x)-\epsilon\right)} \ \mathrm{d}\psi_{\rm B}^\prime\ \mathrm{d}\epsilon\ \mathrm{d}x\
	\end{multline}
	Using Eqs.~(\ref{eq:expression for mass fraction x < x^*, in terms of x}), (\ref{eq:prefactor}) and (\ref{eq:I1 updated to sequential stripping model}), the survival fraction can be computed using Eq.~(\ref{eq:SF expression in terms of mass fraction x < x^* and I_A}).

	\section{Computing the mass fraction of the NFW minihalo below the normalized crossover radius}\label{app:Computing mass fraction below crossover radius}
	In this section, we compute the expression for the mass fraction of the NFW minihalo in the range $x<\min[x^*,1]$. We start with Eq.~(\ref{eq:expression for mass fraction x < x^*, in terms of r}).
	\begin{equation}\label{eq:expression for mass fraction x < x^*, in terms of r - repeated}
		\text{Mass fraction}_{x < \min[x^*,1]} = \frac{1}{M_{\rm vir}} \int\limits_{r=0}^{\min[r^*, r_{\rm vir}]} \frac{\rho_{\rm s}}{\frac{r}{r_{\rm s}}\left(1 + \frac{r}{r_{\rm s}}\right)^2} \times 4\pi r^2\mathrm{d}r
	\end{equation}
	Making the substitution $r=xr_{\rm vir}$ and $c=\frac{r_{\rm vir}}{r_{\rm s}}$ in Eq.~(\ref{eq:expression for mass fraction x < x^*, in terms of r - repeated})
	\begin{equation}\label{eq:expression for mass fraction x < x^*, in terms of r - repeated - dimensionless}
		\text{Mass fraction}_{x < \min[x^*,1]} = \frac{4\pi\rho_{\rm s}r_{\rm vir}^3}{M_{\rm vir}} \int\limits_{x=0}^{\min[x^*, 1]} \frac{1}{cx\left(1 + cx\right)^2} x^2\mathrm{d}x
	\end{equation}
	where $x^* \equiv \frac{r^*}{r_{\rm vir}}$.
	
	The mass enclosed within a sphere of radius $r$ for an NFW density profile is Ref.~\cite[equation 2.66]{binneyTremaine}
	\begin{equation}\label{eq:enclosed mass for NFW}
		M_{\rm enc}(r) = 4\pi\rho_{\rm s}r_{\rm s}^3\left[ \ln\left(1+\frac{r}{r_{\rm s}}\right) +\frac{\frac{r}{r_{\rm s}}}{1+\frac{r}{r_{\rm s}}} \right]
	\end{equation}
	Making the substitutions $\frac{r}{r_{\rm s}} = \frac{xr_{\rm vir}}{r_{\rm s}} = cx$ and $r_{\rm s} = \frac{r_{\rm vir}}{c}$ in Eq.~(\ref{eq:enclosed mass for NFW})
	\begin{equation}\label{eq:Menc(x)}
		M_{\rm enc}(x) = \frac{4\pi\rho_{\rm s}r_{\rm vir}^3}{c^3}\left[ \ln(1+cx) - \frac{cx}{1+cx} \right]
	\end{equation}
	The virial mass $M_{\rm vir}$ is defined as that mass contained within the virial radius ($x=1$). Thus
	\begin{align}\label{eq:M_vir}
		M_{\rm vir} &\equiv M_{\rm enc}(x=1)\nonumber\\
		&= \frac{4\pi\rho_{\rm s}r_{\rm vir}^3}{c^3}\left[ \ln(1+c) - \frac{c}{1+c} \right]\nonumber\\
		&= 4\pi\rho_{\rm s}r_{\rm vir}^3 \frac{f_{\rm NFW}(c)}{c^3}
	\end{align}
	where $f_{\rm NFW}(c) \equiv \ln(1+c) - \frac{c}{1+c}$
	
	Substituting for $M_{\rm vir}$ from Eq.~(\ref{eq:M_vir}) in Eq.~(\ref{eq:expression for mass fraction x < x^*, in terms of r - repeated - dimensionless})
	\begin{equation}\label{eq:expression for mass fraction x < x^*, in terms of x - repeated}
		\text{Mass fraction}_{x < \min[x^*,1]} = \frac{c^2}{f_{\rm NFW}\cg{(c)}} \int\limits_{x=0}^{\min[x^*, 1]} \frac{x}{\left(1 + cx\right)^2} \mathrm{d}x
	\end{equation}
	\section{Computing the expression for the normalized relative potential of an NFW minihalo in the region 
		$x>x^*$ using the sequential stripping model}
	\label{apppsiforxgreaterthanxstarsequentialstrippingmodel}
	For the sequential stripping model, in the region $x>x^*$ where there is complete mass loss, we start by taking the outermost shell to infinity and then the next outermost shell, and so on. Thus, in the region $x>x^*$, when a dark matter particle is taken to infinity, it doesn't encounter a net force from dark matter particles in outer shells. So, the enclosed mass that the dark matter particle sees remains fixed. The Newtonian gravitational potential is Ref.~\cite[equation (2.67)]{binneyTremaine}
	\begin{equation}\label{eq:Phi(r) fundamental equation}
		\Phi(r) = -G \int\limits_{r^\prime=r}^{\infty} \frac{M_{\rm enc}(r^\prime)}{r^{\prime^2}}\mathrm{d}r^\prime
	\end{equation}
	Making the substitution $r^\prime = x^\prime r_{\rm vir}$, Eq.~(\ref{eq:Phi(r) fundamental equation}) becomes
	\begin{equation}\label{eq:Phi(x) intermediate 3}
		\Phi(x) = -\frac{G}{r_{\rm vir}} \int\limits_{x^\prime=x}^{\infty} \frac{M_{\rm enc}(x^\prime)}{x^{\prime^2}}\mathrm{d}x^\prime
	\end{equation}
	where $x \equiv \frac{r}{r_{\rm vir}}$
	
	But the enclosed mass seen by the dark matter particle is always $M_{\rm enc}(x)$, even as it is taken to infinity. Thus
	\begin{align}\label{eq:Phi(x) intermediate 4}
		\Phi(x) &= -\frac{G}{r_{\rm vir}} M_{\rm enc}(x) \int\limits_{x^\prime=x}^{\infty} \frac{1}{x^{\prime^2}}\mathrm{d}x^\prime \nonumber\\
		&= -\frac{G}{r_{\rm vir}} M_{\rm enc}(x) \frac{1}{x}\nonumber\\
		&= -\Psi_0 \frac{M_{\rm enc}(x)}{M_{\rm vir}} \frac{1}{x}
	\end{align}
	where $\Psi_0 \equiv \frac{GM_{\rm vir}}{r_{\rm vir}}$.
	
	Substituting for $M_{\rm enc}(x)$ from Eq.~(\ref{eq:Menc(x)}) and $M_{\rm vir}$ from Eq.~(\ref{eq:M_vir}), Eq.~(\ref{eq:Phi(x) intermediate 4}) becomes
	\begin{align}\label{eq:Phi(x) intermediate 5}
		\Phi(x) &= -\Psi_0 \frac{1}{f_{\rm NFW}(c)} \left[ \ln(1+cx) - \frac{cx}{1+cx} \right] \frac{1}{x} \nonumber\\
		&= -\Psi_0 \frac{1}{f_{\rm NFW}(c)} \left[ \frac{\ln(1+cx)}{x} - \frac{c}{1+cx} \right]
	\end{align}
	Let $\Psi \equiv -\Phi$ and $\psi \equiv \frac{\Psi}{\Psi_0}$. $\implies \psi = -\frac{\Phi}{\Psi_0}$. From Eq.~(\ref{eq:Phi(x) intermediate 5}), this implies that the normalized relative potential is
	\begin{align}
		\psi_{\rm B}(x) = \frac{1}{f_{\rm NFW}(c)} \left[ \frac{\ln(1+cx)}{x} - \frac{c}{1+cx} \right] &&, x>x^*
	\end{align}
	\section{Computing the expression for the normalized relative potential of an NFW minihalo in the region 
		$x<x^*$ using the sequential stripping model}
	\label{app:psi(x) for x < x^* - sequential stripping model}
	Here, we utilize the sequential stripping model. For the purposes of computing the relative potential in the region $x<x^*$, we assume that there is no mass loss in the region $x<x^*$ and complete mass has already occurred in the region $x>x^*$, in keeping with the sequential stripping model. From Eq.~(\ref{eq:Phi(x) intermediate 3}), the Newtonian gravitational potential in the region $x<x^*$ is
	\begin{align}\label{eq:Phi(x) intermediate 6}
		\Phi(x) &= -\frac{G}{r_{\rm vir}} \int\limits_{x^\prime=x}^{\infty} \frac{M_{\rm enc}(x^\prime)}{x^{\prime^2}}\mathrm{d}x^\prime \nonumber\\
		&= -\frac{G}{r_{\rm vir}} \left[\ \int\limits_{x^\prime=x}^{x^*} \frac{M_{\rm enc}(x^\prime)}{x^{\prime^2}}\mathrm{d}x^\prime + M_{\rm enc}(x^*)\int\limits_{x^\prime=x^*}^{\infty} \frac{1}{x^{\prime^2}}\mathrm{d}x^\prime \right]
	\end{align}
	because in the region $x>x^*$, the dark matter particle only sees an enclosed mass of $M_{\rm enc}(x^*)$ since all the mass in the region $x>x^*$ is already stripped off.
	
	Substituting for $M_{\rm enc}(x)$ from Eq.~(\ref{eq:Menc(x)}), Eq.~(\ref{eq:Phi(x) intermediate 6}) becomes
	\begin{align*}
		\Phi(x) = -\frac{G}{r_{\rm vir}} \frac{4\pi\rho_{\rm s}r_{\rm vir}^3}{c^3} \biggr[ &\int\limits_{x^\prime=x}^{x^*} \frac{1}{x^{\prime^2}}\left[ \ln(1+cx^\prime) - \frac{cx^\prime}{1+cx^\prime} \right] \mathrm{d}x^\prime \\
		&+ \left[ \ln(1+cx^*) - \frac{cx^*}{1+cx^*} \right] \frac{1}{x^*} \biggr]
	\end{align*}
	\begin{align*}
		\Phi(x) = -\frac{G}{r_{\rm vir}} \frac{4\pi\rho_{\rm s}r_{\rm vir}^3}{c^3} \biggr[&\left( \frac{\ln(1+cx)}{x} - \frac{\ln(1+cx^*)}{x^*} \right)\\\
		&+ \left( \frac{\ln(1+cx^*)}{x^*} - \frac{c}{1+cx^*} \right) \biggr]
	\end{align*}
	\begin{equation}\label{eq:Phi(x) intermediate 7}
		\Phi(x) = -\Psi_0 \frac{4\pi\rho_{\rm s}r_{\rm vir}^3}{M_{\rm vir}c^3} \left[ \frac{\ln(1+cx)}{x} - \frac{c}{1+cx^*} \right]
	\end{equation}
	Substituting for $M_{\rm vir}$ from Eq.~(\ref{eq:M_vir}), Eq.~(\ref{eq:Phi(x) intermediate 7}) becomes
	\begin{equation}\label{eq:Phi(x) intermediate 8}
		\Phi(x) = -\Psi_0 \frac{1}{f_{\rm NFW}(c)} \left[ \frac{\ln(1+cx)}{x} - \frac{c}{1+cx^*} \right]
	\end{equation}
	Let $\Psi \equiv -\Phi$ and $\psi \equiv \frac{\Psi}{\Psi_0}$. $\implies \psi = -\frac{\Phi}{\Psi_0}$. From Eq.~(\ref{eq:Phi(x) intermediate 8}), this implies that the normalized relative potential is
	\begin{align}
		\psi_{\rm A}(x) = \frac{1}{f_{\rm NFW}\cg{(c)}} \left[ \frac{\ln(1+cx)}{x} - \frac{c}{1+cx^*} \right] &&, x<x^*
	\end{align}

	\section{Computing the expression for the normalized relative potential of the Hernquist density profile using the sequential stripping model}
	\label{app:psi for Hernquist minihalo}
	For a Hernquist profile, the mass enclosed within radius $r$ is Ref.~\cite[equation 2.66]{binneyTremaine}
	\begin{equation}\label{eq:enclosed mass for Hernquist minihalo as a functino of r}
		M_{\rm enc}(r) = 2\pi\rho_{\rm s}r_{\rm s}^3 \frac{\left(\frac{r}{r_{\rm s}}\right)^2}{\left(1+\frac{r}{r_{\rm s}}\right)^2}
	\end{equation}
	Using Eqs. (\ref{eq:x definition}) and (\ref{eq:concentration defintion}), Eq.~(\ref{eq:enclosed mass for Hernquist minihalo as a functino of r}) can be rewritten as
	\begin{equation}\label{eq:enclosed mass for Hernquist minihalo as a functino of x}
		M_{\rm enc}(x) = \frac{2\pi\rho_{\rm s}r_{\rm vir}^3}{c} \frac{x^2}{\left(1+cx\right)^2}
	\end{equation}
	The virial mass is, by definition
	\begin{equation}\label{eq:virial mass of Hernquist minihalo}
		M_{\rm vir} \equiv M_{\rm enc}(x=1) = \frac{2\pi\rho_{\rm s}r_{\rm vir}^3}{c} \frac{1}{(1+c)^2}
	\end{equation}
	From Eqs.~(\ref{eq:enclosed mass for Hernquist minihalo as a functino of x}) and (\ref{eq:virial mass of Hernquist minihalo}), it follows that
	\begin{equation}\label{eq:enclosed mass for Hernquist minihalo as a functino of x, part 2}
		M_{\rm enc}(x) = M_{\rm vir} (1+c)^2 \frac{x^2}{\left(1+cx\right)^2}
	\end{equation}
	According to Eq.~(\ref{eq:Phi(x) intermediate 3}),
	\begin{equation}\label{eq:Phi(x) intermediate 9}
		\Phi(x) = -\frac{G}{r_{\rm vir}} \int\limits_{x^\prime=x}^{\infty} \frac{M_{\rm enc}(x^\prime)}{x^{\prime^2}}\mathrm{d}x^\prime
	\end{equation}
	The normalized crossover radius $x^*$ is defined by Eq.~(\ref{eq:cross over radius condition}). We now look at two cases.
	
	\noindent\underline{Case $x>x^*$}:
	
	In the sequential stripping model, when computing the normalized relative potential $\psi$ at normalized radius $x$, we assume that all shells outward from $x$ have already been stripped off. Thus, as the dark matter particle is taken from $x$ to infinity, the enclosed mass is always $M_{\rm enc}(x)$. Thus Eq.~(\ref{eq:Phi(x) intermediate 9}) becomes
	\begin{equation}\label{eq:Phi(x) intermediate 10}
		\Phi(x) = -\frac{G}{r_{\rm vir}} M_{\rm enc}(x) \frac{1}{x}
	\end{equation}
	\begin{equation}\label{eq:Phi(x) intermediate 11}
		\Phi(x) = -\Psi_0 (1+c)^2\frac{x}{\left(1+cx\right)^2}
	\end{equation}
	where $\Psi_0 \equiv \frac{GM_{\rm vir}}{r_{\rm vir}}$.
	
	Defining $\Psi \equiv -\Phi$, and $\psi \equiv \frac{\Psi}{\Psi_0} = -\frac{\Phi}{\Psi_0}$, we compute the normalized relative potential $\psi$ as
	\begin{align}\label{eq:psi for x>x^* for a Hernquist profile}
		\psi_{\rm B}(x) = (1+c)^2\frac{x}{\left(1+cx\right)^2} &&, x>x^*
	\end{align}
	\noindent\underline{Case $x<x^*$}:
	
	We assume complete mass loss in the region $x>x^*$. In the sequential stripping model, when computing $\psi(x)$, we assume all shells for which $x>x^*$ have already been stripped off. We also assume no shell stripping has occurred in the range $[x,x^*]$. Thus, Eq.~(\ref{eq:Phi(x) intermediate 9}) becomes
	\begin{equation}\label{eq:Phi(x) intermediate 12}
		\Phi(x) = -\frac{G}{r_{\rm vir}} \left[\ \ \int\limits_{x^\prime=x}^{x^*} \frac{M_{\rm enc}(x^\prime)}{x^{\prime^2}}\mathrm{d}x^\prime + \int\limits_{x^\prime=x^*}^{\infty} \frac{M_{\rm enc}(x^*)}{x^{\prime^2}}\mathrm{d}x^\prime \right]
	\end{equation}
	Substituting for $M_{\rm enc}$ from Eq.~(\ref{eq:enclosed mass for Hernquist minihalo as a functino of x, part 2}), Eq.~(\ref{eq:Phi(x) intermediate 12}) becomes
	\begin{equation}\label{eq:Phi(x) intermediate 13}
		\Phi(x) = -\Psi_0 (1+c)^2 \left[ \frac{x^*-x}{(1+cx^*)(1+cx)} + \frac{x^*}{(1+cx^*)^2} \right]
	\end{equation}
	Again, by definition, $\psi = -\frac{\Phi}{\Psi_0}$. We can then compute the normalized relative potential $\psi$ as
	\begin{multline}\label{eq:psi for x<x^* for a Hernquist profile}
		\psi_{\rm A}(x) = (1+c)^2 \left[ \frac{x^*-x}{(1+cx^*)(1+cx)} + \frac{x^*}{(1+cx^*)^2} \right] \\
		, x<x^*
	\end{multline}
	\section{Evaluating the expression for $\alpha^2$ and $\gamma$ for a Hernquist minihalo}
	\label{app:alpha squared and beta for Herquist minihalo}
	\noindent\underline{$\alpha^2$}:
	
	For a spherically symmetric density profile like the Hernquist profile (S2023)
	\begin{equation}\label{eq:alpha squared for Hernquist profile part 1}
		\alpha^2 = \frac{1}{M_{\rm vir}r_{\rm vir}^2}\ \int\limits_{r=0}^{r_{\rm vir}} \mathrm{d}^3\vec{r}\ r^2\rho(r)
	\end{equation}
	Since the density profile is spherically symmetric, Eq.~(\ref{eq:alpha squared for Hernquist profile part 1}) becomes
	\begin{equation}\label{eq:alpha squared for Hernquist profile part 2}
		\alpha^2 = \frac{4\pi}{M_{\rm vir}r_{vir}^2} \int\limits_{r=0}^{r_{\rm vir}} \rho(r)r^4\mathrm{d}r
	\end{equation}
	Instead of $r$, writing the variable of integration as $x$, Eq.(\ref{eq:alpha squared for Hernquist profile part 2}) can be written as
	\begin{equation}\label{eq:alpha squared for Hernquist profile part 3}
		\alpha^2 = \frac{4\pi r_{\rm vir}^3}{M_{\rm vir}}\ \int\limits_{x=0}^1 \rho(x)x^4\mathrm{d}x
	\end{equation}
	Substituting for $\rho(x)$ from Eq.~(\ref{eq:Hernquist profile in terms of x}), we can then compute the one-dimensional integral in Eq.~(\ref{eq:alpha squared for Hernquist profile part 3}). Thus, Eq.~(\ref{eq:alpha squared for Hernquist profile part 3}) becomes
	\begin{equation}\label{eq:alpha squared for Hernquist profile part 4}
		\alpha^2 = \frac{4\pi\rho_{\rm s}r_{\rm vir}^3}{M_{\rm vir}} \left[ \frac{\frac{c(6+9c+2c^2)}{(1+c)^2} - 6\ln(1+c)}{2c^5} \right]
	\end{equation}
	Substituting for $M_{\rm vir}$ from Eq.~(\ref{eq:virial mass of Hernquist minihalo}), Eq.~(\ref{eq:alpha squared for Hernquist profile part 4}) can be written as
	\begin{equation}\label{eq:alpha squared for Hernquist profile part 5}
		\alpha^2 = \frac{c(6+9c+2c^2) - 6(1+c)^2\ln(1+c)}{c^4}
	\end{equation}
	\noindent\underline{$\gamma$}:
	
	For a spherically symmetric density profile like the Hernquist profile (K2021)
	\begin{equation}\label{eq:beta for Hernquist profile part 1}
		\gamma = \frac{4\pi r_{\rm vir}}{M_{\rm vir}^2}\ \int\limits_{r=0}^{r_{\rm vir}} M_{\rm enc}(r)\rho(r)r\mathrm{d}r
	\end{equation}
	Changing the variable of integration from $r$ to $x$, Eq.~(\ref{eq:beta for Hernquist profile part 1}) becomes
	\begin{equation}\label{eq:beta for Hernquist profile part 2}
		\gamma = \frac{4\pi r_{\rm vir}^3}{M_{\rm vir}^2}\ \int\limits_{x=0}^1 M_{\rm enc}(x)\rho(x)x\mathrm{d}x
	\end{equation}
	Substituting for $M_{\rm enc}(x)$ from Eq.~(\ref{eq:enclosed mass for Hernquist minihalo as a functino of x, part 2}), $\rho(x)$ from Eq.~(\ref{eq:Hernquist profile in terms of x}) and $M_{\rm vir}$ from Eq.~(\ref{eq:virial mass of Hernquist minihalo}), the one dimensional integral in Eq.~(\ref{eq:beta for Hernquist profile part 2}) can be evaluated and Eq.~(\ref{eq:beta for Hernquist profile part 2}) can be succinctly written as
	\begin{equation}\label{eq:beta for Hernquist profile part 3}
		\gamma = \frac{4+c}{6}
	\end{equation}

	\section{Computing expressions for the disrupted minihalo's parameters}
	\label{app:Computing expressions for the disrupted minihalo's parameters}
	We now compute each of the terms in Eq.~(\ref{eq:mass condition for transition from NFW minihalo to first-generation minihalo}). For an NFW profile, the mass enclosed within radius $r$ is (Ref.~\cite[Eq. (2.66)]{binneyTremaine})
	\begin{equation}\label{eq:enclosed mass for NFW profile}
		M_{\text {enc,s }}(r)=4 \pi \rho_{\rm s} r_{\rm s}^3\left[\ln \left(1+\frac{r}{r_{\rm s}}\right)-\frac{\frac{r}{r_{\rm s}}}{1+\frac{r}{r_{\rm s}}}\right]
	\end{equation}
	Now,
	\begin{equation}\label{eq:frac{r}{r_s}}
		\frac{r}{r_{\rm s}}=\frac{r}{r_{\rm vir, s}} \frac{r_{\text {vir,s}}}{r_{\rm s}}=x_{\rm s} c_{\rm s}
	\end{equation}
	where
	\begin{equation}\label{eq:c_s defintion}
		c_{\rm s} \equiv \frac{r_{\rm vir,s}}{r_{\rm s}}
	\end{equation}
	Substituting Eq.~(\ref{eq:frac{r}{r_s}}) in Eq.~(\ref{eq:enclosed mass for NFW profile})
	\begin{align}
		M_{\text {enc,s }}\left(x_{\rm s}\right)&=4 \pi \rho_{\rm s} r_{\rm s}^3\left[\ln \left(1+c_{\rm s} x_{\rm s}\right)-\frac{c_{\rm s} x_{\rm s}}{1+c_{\rm s} x_{\rm s}}\right] \nonumber\\
		&=4 \pi \rho_{\rm s} r_{\rm s}^3 f_{\rm NFW}(c_{\rm s} x_{\rm s})
	\end{align}
	\begin{align}\label{eq:mass enclosed within crossover radius for NFW profile}
		\Rightarrow M_{\text {enc,s }}\left(x_{\rm s}^*\right)&=4 \pi \rho_{\rm s} r_{\rm s}^3f_{\rm NFW}(c_{\rm s} x_{\rm s}^*)
	\end{align}
	Next, prefactor $\times I_{\rm A}$ from Eqs.~(\ref{eq:prefactor}) and (\ref{eq:I1 updated to sequential stripping model}), computes the mass loss fraction between $x=0$ and $x=\min \left[x^{*}, 1\right]$. We can convert a mass loss fraction term to a mass loss term by multiplying with $M_{\rm vir}$. Thus,
	\begin{multline}\label{eq:partial mass loss within crossover radius for NFW minihalo}
		\Delta M_{x_{\rm s}=0 \rightarrow x_{\rm s}^*}=M_{\rm vir} \times \frac{4 \pi c_{\rm s}^3}{f_{\rm NFW}(c_{\rm s})} \int\limits_{x=0}^{x_{\rm s}^*} \int\limits_{\epsilon=0}^{\vert\Delta\epsilon(x)\vert} \  \int\limits_{\psi_{\rm B}^\prime=0}^\epsilon x^2\frac{1}{\sqrt{8}\pi^2}\\ \frac{1}{\sqrt{\epsilon - \psi_{\rm B}^\prime}} \frac{\mathrm{d}^2\varrho}{\mathrm{d}\psi_{\rm B}^{\prime^2}}\left(x^\prime\left(\psi_{\rm B}^\prime\right)\right) \sqrt{2\left(\psi_{\rm A}(x)-\epsilon\right)} \ \mathrm{d}\psi_{\rm B}^\prime\ \mathrm{d}\epsilon\ \mathrm{d}x\
	\end{multline}
	Substituting for $M_{\rm vir}$ from Eq.~(\ref{eq:M_vir}) in Eq.~(\ref{eq:partial mass loss within crossover radius for NFW minihalo})
	\begin{align}\label{eq:partial mass loss within crossover radius for NFW minihalo, intermediate 2}
		\Delta M_{x_{\rm s}=0 \rightarrow x_{\rm s}^*} &= 16\pi^2\rho_{\rm s}r_{\rm s}^3c_{\rm s}^3 \int\limits_{x=0}^{x_{\rm s}^*} \int\limits_{\epsilon=0}^{\vert\Delta\epsilon(x)\vert} \  \int\limits_{\psi_{\rm B}^\prime=0}^\epsilon\nonumber\\
		&\ \ \ \ \ \ \ \ \ \ x^2\frac{1}{\sqrt{8}\pi^2} \frac{1}{\sqrt{\epsilon - \psi_{\rm B}^\prime}} \frac{\mathrm{d}^2\varrho}{\mathrm{d}\psi_{\rm B}^{\prime^2}}\left(x^\prime\left(\psi_{\rm B}^\prime\right)\right) \nonumber\\
		&\ \ \ \ \ \ \ \ \ \ \sqrt{2\left(\psi_{\rm A}(x)-\epsilon\right)} \ \mathrm{d}\psi_{\rm B}^\prime\ \mathrm{d}\epsilon\ \mathrm{d}x\nonumber\\
		&= 16\pi^2\rho_{\rm s}r_{\rm s}^3c_{\rm s}^3 \times I_{\rm s}
	\end{align}
	where
	\begin{multline}\label{eq:I_s definition}
		I_{\rm s} = \int\limits_{x=0}^{x_{\rm s}^*} \int\limits_{\epsilon=0}^{\vert\Delta\epsilon(x)\vert} \  \int\limits_{\psi_{\rm B}^\prime=0}^\epsilon x^2\frac{1}{\sqrt{8}\pi^2} \frac{1}{\sqrt{\epsilon - \psi_{\rm B}^\prime}} \frac{\mathrm{d}^2\varrho}{\mathrm{d}\psi_{\rm B}^{\prime^2}}\left(x^\prime\left(\psi_{\rm B}^\prime\right)\right)\\
		\sqrt{2\left(\psi_{\rm A}(x)-\epsilon\right)} \ \mathrm{d}\psi_{\rm B}^\prime\ \mathrm{d}\epsilon\ \mathrm{d}x
	\end{multline}
	Next, in general, for a broken power law of the form 
	given by Eq. (\ref{eq:broken power law})
	the total enclosed mass is
	\begin{align}\label{eq:total mass of broken power law profile with k = 2 + Delta}
		\lim _{x_1 \rightarrow \infty} M_{\rm enc,1}\left(x_1\right) &= \lim _{r \rightarrow \infty} M_{\text {enc,1}}(r) \nonumber\\
		& =\int\limits_{r=0}^{\infty} \rho_{k=2+\Delta}(r) \times 4 \pi r^2 d r \nonumber\\
		& =\int\limits_{r=0}^{\infty} \frac{\rho_1}{\frac{r}{r_1}\left(1+\frac{r}{r_1}\right)^{2+\Delta}} 4 \pi r^2 d r \nonumber\\
		& =\frac{4 \pi \rho_1 r_1^3}{\Delta+\Delta^2}
	\end{align}
	Finally, substituting Eqs.~(\ref{eq:mass enclosed within crossover radius for NFW profile}), (\ref{eq:partial mass loss within crossover radius for NFW minihalo, intermediate 2}) and (\ref{eq:total mass of broken power law profile with k = 2 + Delta}) in Eq.~(\ref{eq:mass condition for transition from NFW minihalo to first-generation minihalo})
	
	\begin{align}\label{eq:mass condition for transition from NFW minihalo to first-generation minihalo substituted}
		4 \pi \rho_{\rm s} r_{\rm s}^3 f_{\rm NFW}(c_{\rm s} x_{\rm s}^*)-16 \pi^2 \rho_{\rm s} r_{\rm s}^3 c_{\rm s}^3 I_{\rm s} =\frac{4 \pi \rho_1 r_1^3}{\Delta+\Delta^2}
	\end{align}
	
	Dividing Eq.~(\ref{eq:mass condition for transition from NFW minihalo to first-generation minihalo substituted}) by $4 \pi \rho_{\rm s} r_{\rm s}^3$
	\begin{equation}\label{eq:mass condition for transition from NFW minihalo to first-generation minihalo substituted, intermediate 2}
		f_{\rm NFW}(c_{\rm s} x_{\rm s}^*)-4 \pi c_{\rm s}^3 I_{\rm s}=\frac{1}{\Delta + \Delta^2}\left(\frac{\rho_1 r_1}{\rho_{\rm s} r_{\rm s}}\right)\left(\frac{r_1}{r_{\rm s}}\right)^2
	\end{equation}
	
	Substituting Eq.~(\ref{eq:small radius condition}) in Eq.~(\ref{eq:mass condition for transition from NFW minihalo to first-generation minihalo substituted, intermediate 2}), we can then compute $r_1$ in terms of $r_{\rm s}$ as follows
	\begin{align}\label{eq:r_1 final expression - broken power law k=3.2}
		r_1&=r_{\rm s} \sqrt{(\Delta + \Delta^2)\left[f_{\rm NFW}(c_{\rm s} x_{\rm s}^*)-4 \pi c_{\rm s}^3 I_{\rm s}\right]} \nonumber\\
		&=r_{\rm s} \times R_{\rm s}
	\end{align}
	where
	\begin{equation}\label{eq:r_s defintion - broken power law k=3.2}
		R_{\rm s} \equiv \sqrt{(\Delta + \Delta^2)\left[f_{\rm NFW}(c_{\rm s} x_{\rm s}^*)-4 \pi c_{\rm s}^3 I_{\rm s}\right]}
	\end{equation}
	Substituting Eq.~(\ref{eq:r_1 final expression - broken power law k=3.2}) in Eq.~(\ref{eq:small radius condition}), we get
	\begin{equation}\label{eq:rho_1 final condition}
		\rho_1 = \frac{\rho_{\rm s}}{R_{\rm s}}
	\end{equation}

	\section{Computing the expression for survival fraction, incorporating relaxation}
	\label{app:Computing the expression for survival fraction with relaxation}
	Here, we compute the expression for the survival fraction, assuming that the remnant minihalo following a stellar encounter (with an NFW minihalo) relaxes to a Hernquist profile. Our region of interest for calculating the survival fraction is the physical virial radius of the unperturbed NFW minihalo. Thus, the survival fraction is the ratio of the mass enclosed by the region of interest for the relaxed Hernquist profile to the mass enclosed by the same region of interest for the unperturbed NFW profile which is just the virial mass $M_{\text{vir,s}}$ of the NFW minihalo.
	
	\begin{equation}\label{eq:survival fraction incorporating relaxation - appendix version}
		\text {SF} \equiv \frac{M_{\text {enc,1}}\left(x_1^{r_{\text{vir,s}}}\right)}{M_{\text{vir,s}}}
	\end{equation}
	Here, $x_1^{r_{\text{vir,s}}}$ is the physical virial radius of the unperturbed NFW minihalo expressed in the ``local" normalized radial distance variable of the relaxed Hernquist minihalo.
	
	\begin{align}
		x_1^{r_{\text{vir,s}}} &\equiv \frac{r_{\rm vir,s}}{r_{\rm vir,1}} \\
		& =\frac{c_{\rm s} r_{\rm s}}{c_1 r_1} \nonumber\\
		& =\frac{c_{\rm s}}{c_1} \frac{1}{R_{\rm s}} 
	\end{align}
	where $r_{\rm vir,s}$ and $r_{\rm vir,1}$ are the physical virial radii of the unperturbed NFW minihalo and the relaxed Hernquist minihalo, respectively. $c_{\rm s}$ and $c_1$ are the concentrations of the NFW and Hernquist minihalos respectively.
	According to  Ref.~\cite[Eq.(2.66)]{binneyTremaine}
	\begin{align}
		M_{\rm enc,1}(r) &= 2 \pi \rho_1  r_1^3 \frac{\left(r / r_1\right)^2}{\left(1+r / r_1\right)^2} \\
		\implies M_{\rm enc,1}\left(x_1\right) &= 2 \pi \rho_1 r_1^3 \frac{\left(c_1 x_1\right)^2}{\left(1+c_1 x_1\right)^2}
	\end{align}
	where $x_1 \equiv \frac{r}{r_{\rm vir, 1}}$.
	Thus,
	
	\begin{align}\label{eq:mass enclosed within physical virial radius (of NFW minihalo) for a Hernquist profile}
		M_{\rm enc,1}\left(x_1^{r_{\rm vir,s}}\right) &= 2 \pi \rho_1 r_1^3 \frac{\left(c_1 x_1^{r_{\rm vir,s}}\right)^2}{\left(1+c_1 x_1^{r_{\rm vir,s}}\right)^2} \nonumber\\
		& =2 \pi \rho_1 r_1^3 f_{\rm Hern}(c_1 x_1^{r_{\rm vir,s}})
	\end{align}
	Next, $M_{\rm vir,s}$ is given by Eq.~(\ref{eq:M_vir}) but it can be rewritten as
	
	\begin{equation}\label{eq:virial mass of NFW profile using r_s}
		M_{\rm vir, s} = 4 \pi \rho_{\rm s} r_{\rm s}^3 f_{\rm NFW}(c_{\rm s})
	\end{equation}
	Substituting Eqs.~(\ref{eq:mass enclosed within physical virial radius (of NFW minihalo) for a Hernquist profile}) and (\ref{eq:virial mass of NFW profile using r_s}) in Eq.~(\ref{eq:survival fraction incorporating relaxation - appendix version})
	\begin{align}\label{eq:survival fraction incorporating relaxation; pref-final expression}
		\text{SF} &=\frac{2 \pi \rho_1 r_1^3 f_{\rm Hern}(c_1 x_1^{r_{\rm vir,s}})}{4 \pi \rho_{\rm s} r_{\rm s}^3 f_{\rm NFW}(c_{\rm s})} \nonumber\\
		& =\frac{1}{2}\left(\frac{\rho_1 r_1}{\rho_{\rm s} r_{\rm s}}\right)\left(\frac{r_1}{r_{\rm s}}\right)^2 \frac{f_{\rm Hern}(c_1 x_1^{r_{\rm vir,s}})}{f_{\rm NFW}(c_{\rm s})}
	\end{align}
	Substituting Eqs.~(\ref{eq:small radius condition}) and (\ref{eq:r_1 final expression - broken power law k=3.2}) in Eq.~(\ref{eq:survival fraction incorporating relaxation; pref-final expression})
	
	\begin{equation}
		\text{SF} =\frac{1}{2} R_{\rm s}^2 \frac{f_{\rm Hern}(c_1 x_1^{r_{\rm vir,s}})}{f_{\rm NFW}(c_{\rm s})}
	\end{equation}
	\section{Evaluating mass loss under multiple stellar encounters of an NFW minihalo}
	\label{app:multiple stellar encounters}
	\subsection{Computing the concentration of a Hernquist minihalo given its scale density}\label{app:computing the concentration of Hernquist minihalo, given its scale density}
	We start with the definition of the virial radius of the Hernquist minihalo, analogous to Eq.~(\ref{eq:Virial radius condition}). This leads us to an equation similar to Eq.~(\ref{eq:Virial radius condition, intermediate 2}) but for the first-generation Hernquist minihalo. Thus, we have
	
	\begin{equation}\label{eq:virial radius condition intermediate for Hernquist profile}
		\int\limits_{x_1=0}^1 \rho_{\text {Hern }}\left(x_1\right) x_1^2 d x_1=\frac{200}{3} \rho_{\text {crit }}
	\end{equation}
	where
	\begin{equation}
		x_1 \equiv \frac{r}{r_{\rm vir,1}}
	\end{equation}
	and $r_{\rm vir,1}$ is the virial radius of the first-generation Hernquist minihalo. Substituting Eq.~(\ref{eq:Hernquist profile in terms of x}) in Eq.(\ref{eq:virial radius condition intermediate for Hernquist profile}) and performing the integral with respect to $x_1$ in the L.H.S. of Eq.(\ref{eq:virial radius condition intermediate for Hernquist profile}), we get
	
	\begin{equation}\label{eq:relating concentration and scale radius for Hernquist profile, appendix version}
		\frac{1}{2 c_1\left(1+c_1\right)^2}=\frac{200}{3}\frac{\rho_{\text {crit}}}{\rho_1}
	\end{equation}
	\subsection{Evaluating the normalized virial radius of the unperturbed NFW minihalo, expressed in the local variable of the first-generation Hernquist minihalo}\label{app:normalized radius of NFW minihalo, expresssed in terms of local variable of first-generation minihalo}
	Here, we compute $x_1^{r_{\rm vir,s}}$, the normalized virial radius of the unperturbed NFW minihalo, expressed in the local variable of the first-generation Hernquist minihalo, as follows
	
	\begin{align}
		x_1^{r_{\text {vir,s}}} & \equiv \frac{r_{\text {vir,s}}}{r_{\text {vir,1}}} \nonumber\\
		& =\frac{c_{\rm s} r_{\rm s}}{c_1 r_1} \nonumber\\
		& =\frac{c_{\rm s}}{c_1} \frac{1}{R_{\rm s}}
	\end{align}
	\subsection{Computing the ratio of scale radii of the $(n+1)^{\text{th}}$ and $n^{\text{th}}$ generation minihalos. Also, computing the scale density of the $(n+1)^{\text{th}}$ generation minihalo}\label{app:rn+1_by_rn and rho_n+1 derivation}
	We start with Eq.~(\ref{eq:mass condition for transition from n^th to (n+1)^th generation Hernquist minihalo}) and evaluate each of the three terms in this equation. Both the $n^{\rm th}$ and $(n+1)^{\rm th}$ generation minihalos have a Hernquist density profile. Firstly, adapting Eq.~(\ref{eq:enclosed mass for Hernquist minihalo as a functino of r}), the mass enclosed by the $n^{\rm th}$ generation Hernquist minihalo is
	
	\begin{equation}\label{eq:enclosed mass for (n-1)th generation Hernquist minihalo as a function of r}
		M_{\text {enc,} n}(r)=2 \pi \rho_n r_n^3 \frac{\left(\frac{r}{r_n}\right)^2}{\left(1+\frac{r}{r_n}\right)^2}
	\end{equation}
	But
	
	\begin{equation}\label{eq:r by r_(n-1) expression}
		\frac{r}{r_n}=\frac{r}{r_{\text{vir},n}} \frac{r_{\text{vir},n}}{r_n}=x_n c_n
	\end{equation}
	where $r_{\text{vir},n}$ is the viral radius of the $n^{\text{th}}$ generation minihalo, and
	
	\begin{equation}
		x_n \equiv \frac{r}{r_{\text{vir},n}}
	\end{equation}
	\begin{equation}
		c_n \equiv \frac{r_{\text{vir},n}}{r_n}
	\end{equation}
	Substituting Eq.~(\ref{eq:r by r_(n-1) expression}) in Eq.~(\ref{eq:enclosed mass for (n-1)th generation Hernquist minihalo as a function of r}),
	\begin{align}\label{eq:mass enclosed by the n^th generation Hernquist minihalo as a function of x}
		M_{\text {enc,}n}(x_n)&=2 \pi \rho_n r_n^3 \frac{\left(c_n x_n\right)^2}{\left(1+c_n x_n\right)^2} \nonumber\\
		&= 2 \pi \rho_n r_n^3 f_{\rm Hern}(c_n x_n)
	\end{align}
	When $x_n=x_n^*$, the normalized crossover radius of the $n^{\text{th}}$ generation Hernquist minihalo,
	\begin{equation}\label{eq:mass enclosed within normalized crossover radius for (n-1)th generation minihalo}
		M_{\text {enc,}n}(x_n^*)=2 \pi \rho_n r_n^3 f_{\rm Hern}(c_n x_n^*)
	\end{equation}
	
	Secondly, Eq.~(\ref{eq:Delta M dimensionless}) gives the mass loss between $x=0$ and $x=1$ for a Hernquist (as well as NFW) profile. To evaluate $\Delta M_{x_n=0 \rightarrow x_n^*}$, we need to evaluate the mass loss between $x=0$ and $x=x^*$. Between these limits, Fig.~\ref{fig:psi crossing over deltaEpsilon} tells us that $\min[\vert\Delta\epsilon(x)\vert, \psi(x)] = \vert\Delta\epsilon(x)\vert$. Thus, Eq.~(\ref{eq:Delta M dimensionless}) turns into
	\begin{multline}\label{eq:mass loss between x=0 and x=min[x^*,1]}
		\Delta M_{x_n=0 \rightarrow x_n^*} = 16\pi^2\rho_n r_{\text{vir},n}^3 \int\limits_{x_n=0}^{x_n^*} \mathrm{d}x_n\ x_n^2\\
		\int\limits_{\epsilon=0}^{\vert\Delta\epsilon(x_n)\vert} \mathrm{d}\epsilon\sqrt{2(\psi_{\rm A}(x_n)-\epsilon)}\hat{f}(\epsilon)
	\end{multline}
	Substituting $\hat{f}(\epsilon)$ from Eq.~(\ref{eq:f_hat}) in Eq.~(\ref{eq:mass loss between x=0 and x=min[x^*,1]}),
	\begin{align}\label{eq:partial mass lost inside normalized crossover radius for (n-1)th generation minihalo}
		\Delta M_{x_n=0 \rightarrow x_n^*} &= 16\pi^2\rho_nr_{\text{vir},n}^3  \nonumber\\
		&\mkern-18mu\mkern-18mu\mkern-18mu\mkern-18mu\mkern-18mu \int\limits_{x_n=0}^{x_n^*}\int\limits_{\epsilon=0}^{\vert\Delta\epsilon(x_n)\vert} \int\limits_{\psi_{\rm B}^\prime=0}^\epsilon \frac{1}{\sqrt{8}\pi^2} x_n^2 \sqrt{2(\psi_{\rm A}(x_n)-\epsilon)} \nonumber\\
		&\mkern-18mu\mkern-18mu\mkern-18mu\mkern-18mu\mkern-18mu\frac{1}{\sqrt{\epsilon - \psi_{\rm B}^\prime}} \frac{\mathrm{d}^2\varrho}{\mathrm{d}\psi_{\rm B}^{\prime^2}} \left(x_n^\prime(\psi_{\rm B}^\prime)\right) \mathrm{d}\mathrm{\psi_{\rm B}^\prime} \mathrm{d}\epsilon \mathrm{d}x_n \nonumber\\
		&= 16\pi^2\rho_nc_n^3r_n^3 \times I_n
	\end{align}
	where
	\begin{multline}
		I_n \equiv \int\limits_{x_n=0}^{x_n^*}\int\limits_{\epsilon=0}^{\vert\Delta\epsilon(x_n)\vert} \int\limits_{\psi_{\rm B}^\prime=0}^\epsilon \frac{1}{\sqrt{8}\pi^2} x_n^2 \sqrt{2(\psi_{\rm A}(x_n)-\epsilon)} \\
		\frac{1}{\sqrt{\epsilon - \psi_{\rm B}^\prime}} \frac{\mathrm{d}^2\varrho}{\mathrm{d}\psi_{\rm B}^{\prime^2}} \left(x_n^\prime(\psi_{\rm B}^\prime)\right) \mathrm{d}\mathrm{\psi_{\rm B}^\prime}\mathrm{d}\epsilon \mathrm{d}x_n
	\end{multline}
	Thirdly,
	
	\begin{align}\label{eq:total mass of (n)th generation minihalo}
		\lim _{x_{n+1} \rightarrow \infty} M_{\text{enc},n+1}\left(x_{n+1}\right) & =\lim _{r \rightarrow \infty} M_{\text {enc,}n+1}(r) \nonumber\\
		& =\lim _{r \rightarrow \infty} 2 \pi \rho_{n+1} r_{n+1}^3 \frac{\left(\frac{r}{r_{n+1}}\right)^2}{\left(1+\frac{r}{r_{n+1}}\right)^2} \nonumber\\
		& =2 \pi \rho_{n+1} r_{n+1}^3
	\end{align}
	Substituting Eqs.~(\ref{eq:mass enclosed within normalized crossover radius for (n-1)th generation minihalo}), (\ref{eq:partial mass lost inside normalized crossover radius for (n-1)th generation minihalo}) and (\ref{eq:total mass of (n)th generation minihalo}) in Eq.~(\ref{eq:mass condition for transition from n^th to (n+1)^th generation Hernquist minihalo})
	\begin{equation}\label{eq:mass condition for transition from (n-1)th gneeration Hernquist minihalo to (n)th generation Hernquist minihalo; expressions substituted}
		2 \pi \rho_n r_n^3 f_{\rm Hern}(c_n x_n^*)-16 \pi^2 \rho_n c_n^3 r_n^3 I_n =2 \pi \rho_{n+1} r_{n+1}^3
	\end{equation}
	Dividing Eq.~(\ref{eq:mass condition for transition from (n-1)th gneeration Hernquist minihalo to (n)th generation Hernquist minihalo; expressions substituted}) by $2 \pi \rho_n r_n^3$
	
	\begin{equation}\label{eq:mass condition for transition from (n-1)th gneeration Hernquist minihalo to (n)th generation Hernquist minihalo; expressions substituted and divided throughout by apprpriate quantity}
		f_{\rm Hern}(c_n x_n^*) - 8 \pi c_n^3 I_n = \frac{\rho_{n+1} r_{n+1}}{\rho_n r_n} \left( \frac{r_{n+1}}{r_n} \right)^2
	\end{equation}
	Here too, we assume that at small radii, the $n^{\text{th}}$ and $(n+1)^{\text{th}}$ generation minihalos are indistinguishable from each other. Thus, we arrive at a similar ``small radius condition" as Eq.~(\ref{eq:small radius condition}):
	\begin{equation}\label{eq:small radius condition for (n-1)th and (n)th generation minihalos}
		\rho_n r_n = \rho_{n+1} r_{n+1}
	\end{equation}
	Substituting Eq.~(\ref{eq:small radius condition for (n-1)th and (n)th generation minihalos}) in Eq.~(\ref{eq:mass condition for transition from (n-1)th gneeration Hernquist minihalo to (n)th generation Hernquist minihalo; expressions substituted and divided throughout by apprpriate quantity}), we get the ratio
	\begin{align}\label{eq:r_n by r_(n-1); appendix version}
		\frac{r_{n+1}}{r_n} & =\sqrt{f_{\rm Hern}(c_n x_n^*)-8 \pi c_n^3 I_n} \nonumber\\
		& =R_n
	\end{align}
	where
	\begin{equation}\label{eq:R_(n-1) definition; appendix version}
		R_n \equiv \sqrt{f_{\rm Hern}(c_n x_n^*)-8 \pi c_n^3 I_n}
	\end{equation}
	Substituting Eq.~(\ref{eq:r_n by r_(n-1); appendix version}) in Eq.~(\ref{eq:small radius condition for (n-1)th and (n)th generation minihalos})
	
	\begin{equation}\label{eq:rho_n; appendix version}
		\rho_{n+1} = \frac{\rho_n}{R_n}
	\end{equation}
	\subsection{Computing the survival fraction of the $n^{\text{th}}$ generation minihalo}\label{app:SF_n derivation}
	Here, we compute the survival fraction of the $n^{\text{th}}$ generation Hernquist minihalo, assuming it relaxes to the $(n+1)^{\text{th}}$ generation Hernquist minihalo. Our region of interest for calculating the mass loss is the physical virial radius of the unperturbed NFW minihalo.
	
	We subject the $n^{\text{th}}$ generation Hernquist minihalo to a stellar encounter and let the remnant minihalo relax to an $(n+1)^{\text{th}}$ generation Hernquist minihalo. The survival fraction of the $n^{\text{th}}$ generation minihalo is then given by the ratio of the mass enclosed by the relaxed $(n+1)^{\text{th}}$ generation Hernquist minihalo within the region of interest to the mass enclosed by the unperturbed NFW minihalo within the same region of interest. Thus,
	
	\begin{equation}\label{eq:survival fraction of n^th generation minihalo; definition}
		\text{SF}_n \equiv \frac{M_{\text{enc},n+1}\left(x_{n+1}^{r_{\text{vir,s}}}\right)}{M_{\text{vir,s}}}
	\end{equation}
	where
	
	\begin{equation}
		x_{n+1}^{r_{\text{vir,s}}}=\frac{r_{\text{vir,s}}}{r_{\text{vir,}n+1}}
	\end{equation}
	is the physical virial radius of the unperturbed NFW minihalo expressed in the normalized ``local" radial distance variable of the $(n+1)^{\text{th}}$ generation minihalo. Thus,
	
	\begin{align}
		x_{n+1}^{r_{\rm vir,s}} & =\frac{c_{\rm s} r_{\rm s}}{c_{n+1} r_{n+1}} \nonumber\\
		& =\frac{c_{\rm s}}{c_{n+1}} \frac{1}{\frac{r_{n+1}}{r_n} \frac{r_n}{r_{n-1}} \cdots \frac{r_2}{r_1} \frac{r_1}{r_{\rm s}}} \nonumber\\
		& =\frac{c_{\rm s}}{c_{n+1}} \frac{1}{R_n R_{n-1} \cdots R_1 R_{\rm s}}
	\end{align}
	where
	\begin{equation}
		R_i=\frac{r_{i+1}}{r_i}
	\end{equation}
	Next, adapting Eq.~(\ref{eq:mass enclosed by the n^th generation Hernquist minihalo as a function of x}), the mass enclosed by the $(n+1)^{\text{th}}$ generation Hernquist profile is
	
	\begin{align}\label{eq:mass enclosed by (n+1)^th generation Hernquist minihalo inside the physical virial radius of the NFW minihalo}
		M_{\text {enc,}n+1}\left(x_{n+1}\right) & =2 \pi \rho_{n+1} r_{n+1}^3 f_{\rm Hern}(c_{n+1} x_{n+1}) \nonumber\\
		\implies M_{\text {enc,}n+1}\left(x_{n+1}^{r_{\rm vir,s}}\right) & =2 \pi \rho_{n+1} r_{n+1}^3 f_{\rm Hern}(c_{n+1} x_{n+1}^{r_{\rm vir,s}})
	\end{align}
	Substituting Eqs.~(\ref{eq:mass enclosed by (n+1)^th generation Hernquist minihalo inside the physical virial radius of the NFW minihalo}) and (\ref{eq:virial mass of NFW profile using r_s}) in Eq.~(\ref{eq:survival fraction of n^th generation minihalo; definition})
	
	\begin{align}\label{eq:survival fraction of n^th generation minihalo; intermediate 1}
		\text{SF}_n &= \frac{2 \pi \rho_{n+1} r_{n+1}^3 f_{\rm Hern}(c_{n+1} x_{n+1}^{r_{\rm vir,s}})}{4 \pi \rho_{\rm s} r_{\rm s}^3 f_{\rm NFW}(c_{\rm s})} \nonumber\\
		& =\frac{1}{2} \frac{\rho_{n+1} r_{n+1}}{\rho_{\rm s} r_{\rm s}}\left(\frac{r_{n+1}}{r_{\rm s}}\right)^2 \frac{f_{\rm Hern}(c_{n+1} x_{n+1}^{r_{\rm vir,s}})}{f_{\rm NFW}(c_{\rm s})} 
	\end{align}
	Mandating that all generations of minihalos are indistinguishable at small radii, we have
	
	\begin{equation}\label{eq:small radius condition for NFW and (n+1)^th generation minihalos}
		\rho_{\rm s} r_{\rm s} = \rho_{n+1} r_{n+1}
	\end{equation}
	Moreover,
	
	\begin{align}\label{eq:ratio of scale radii of (n+1)^th generation minihalo to NFW minihalo}
		\frac{r_{n+1}}{r_{\rm s}} & =\frac{r_{n+1}}{r_n} \frac{r_n}{r_{n-1}} \cdots \frac{r_2}{r_1} \frac{r_1}{r_{\rm s}} \nonumber\\
		& =R_n R_{n-1} \cdots R_1 R_{\rm s}
	\end{align}
	Substituting Eqs.~(\ref{eq:small radius condition for NFW and (n+1)^th generation minihalos}) and (\ref{eq:ratio of scale radii of (n+1)^th generation minihalo to NFW minihalo}) in Eq.~(\ref{eq:survival fraction of n^th generation minihalo; intermediate 1})
	
	\begin{equation}
		\text{SF}_n = \frac{1}{2}\left(R_n R_{n-1} \cdots R_1 R_{\rm s}\right)^2 \frac{f_{\rm Hern}(c_{n+1} x_{n+1}^{r_{\rm vir,s}})}{f_{\rm NFW}(c_{\rm s})}
	\end{equation}

\end{document}